\documentclass[a4paper,11pt]{article}
\pdfoutput=1 
             
\usepackage{jheppub} 
                     
\pdfoutput=1
\usepackage{graphicx,array}
\usepackage{color}
\usepackage{setspace}
\usepackage{amstext}
\usepackage{amssymb}
\usepackage{slashed}
\usepackage{makecell}
\usepackage{multirow}
\usepackage{wasysym}
\usepackage{physics}
\usepackage[T1]{fontenc} 
\usepackage{changepage}
\usepackage{arydshln}
\usepackage{mathrsfs} 
\usepackage{comment}
\usepackage{tikz}
\usetikzlibrary{tikzmark}                 

\definecolor{orange}{rgb}{1,0.5,0}
\definecolor{brown}{rgb}{0.59, 0.29, 0.0}

\definecolor{note_fontcolor}{rgb}{0.80078125, 0.80078125, 0.80078125}

\newcommand{\shc}[1]{\textbf{\textcolor{blue}{(#1 --SH)}}}

\def\beq{\begin{equation}}
\def\eeq{\end{equation}}
\def\bea{\begin{eqnarray}}
\def\eea{\end{eqnarray}}

\DeclareMathOperator{\diag}{diag}
\let\originalleft\left
\let\originalright\right
\renewcommand{\left}{\mathopen{}\mathclose\bgroup\originalleft}
\renewcommand{\right}{\aftergroup\egroup\originalright}

\makeatletter
\newcommand*{\Relbarfill@}{\arrowfill@\Relbar\Relbar\Relbar}
\newcommand*{\xeq}[2][]{\ext@arrow 0055\Relbarfill@{#1}{#2}}
\makeatother

\newcommand*\circled[1]{\tikz[baseline=(char.base)]{
    \node[shape=circle,draw,inner sep=1.0pt] (char) {#1};}}

\title{\boldmath Anomaly Inflow and Holography}

\author[a,b,c]{Sungwoo Hong,}
\author[d]{and Gabriele Rigo}

\affiliation[a]{Department of Physics, LEPP, Cornell University, Ithaca, NY 14853, USA}
\affiliation[b]{Department of Physics, The University of Chicago, Chicago, IL 60637 , USA }
\affiliation[c]{Argonne National Laboratory, Lemont, IL 60439, USA}
\affiliation[d]{Department of Physics, Syracuse University, Syracuse, NY 13244, USA}

\abstract{We systematically study the perturbative anomaly inflow by the bulk Chern-Simons (CS) theory in a slice of five-dimensional anti-de Sitter spacetime ($\text{AdS}_5$). 
The introduction of UV and IR 3-branes makes the anomaly story remarkably rich and many interesting aspects can be obtained, including weakly gauging and spontaneous symmetry breaking of the global symmetries of the dual 4D CFT. Our main contribution is to provide a unified and comprehensive discussion of the subject, together with a detailed description of the dual CFT picture for each case. To this end, we employ a gauge-fixed effective action suitable for a holographic study, which allows us to incorporate general UV and IR boundary conditions (BCs). As part of the process, we reproduce many known results in the literature, such as 't Hooft anomaly matching for unbroken symmetry (Neumann IR-BC) and (gauged) Wess-Zumino-Witten (WZW) action for broken symmetry (IR-BC breaks the bulk group $G \to H$). In addition, we show that anomaly matching occurs for ABJ anomalies as well as 't Hooft anomalies, which suggests anomalies inflowed from the bulk CS theory are necessarily free of mixed anomalies with the confining gauge force of the 4D dual CFT. In the case of broken symmetry, we prove that the ``would-be'' Goldstone bosons associated with the weakly gauged symmetry are completely removed by a proper field redefinition, provided the anomaly from the bulk is exactly cancelled by the boundary contribution, hence confirming the standard expectation. Moreover, we present 
a holographic formulation of Witten's argument for the quantization condition for the WZW action, and argue in favor of an alternative way to obtain the same condition using a ``deformed'' theory (different BCs). We work out several examples, including a product group with mixed anomaly, and identify the corresponding dual CFT picture. We consider a fully general case typically arising in the context of dynamical electroweak symmetry breaking. 

}

\begin{document}
\maketitle

\section{Introduction}
\label{sec:Intro }

Fermion anomalies have played an important role in a plethora of different aspects in theoretical physics. The discovery of the one-loop triangle anomaly of Adler, Bell, and Jackiw (ABJ)~\cite{Adler:1969gk, Bell:1969ts} taught us that fermion anomalies not only have direct implications on observable phenomena such as $\pi^0 \to \gamma \gamma$, but also provide strong constraints on consistent quantum field theories. 
That chiral anomalies do not receive any renormalization beyond one-loop (Adler-Bardeen theorem~\cite{Adler:1969er}, see also~\cite{Adler:2004qt} for a review) was then beautifully realized in `t Hooft anomaly matching argument~\cite{tHooft:1980xss} showing that fermion anomalies can yield non-trivial consequences in the spectrum of the infrared (IR) phase of the confining gauge theory. Applications of anomaly matching (together with other sets of techniques) resulted in tremendous success in uncovering phases and dualities of supersymmetric gauge theories~\cite{Seiberg:1994pq, Intriligator:1995au, Peskin:1997qi}. In theories with spontaneously broken global symmetries, the chiral anomalies of the ultraviolet (UV) phase of the theory are maintained in the IR by the (gauged) Wess-Zumino-Witten (WZW) action~\cite{Wess:1971yu, Witten:1983tw}. This also led to an elegant resolution of the puzzle of CP-violation, e.g.~$K^+ K^- \to \pi^+ \pi^- \pi^0$, which is absent in any order in chiral perturbation theory~\cite{Witten:1983tw}. In string theory, the requirement of anomaly cancellation was crucial in arriving at the conclusion that for any supersymmetric theory in 10D with gravity and gauge supermultiplets, the only allowed gauge groups are $SO(32)$ or $E_8 \times E_8$~\cite{Green:1984sg, Adams:2010zy}. 

Fermion anomalies in spacetime dimension $D=2n$ often find their natural description from a higher-dimensional setup. Notably, the form of chiral anomalies compatible with the Wess-Zumino consistency condition~\cite{Wess:1971yu} can be constructed by starting from an Abelian anomaly in $2n+2$ dimensions and subsequently arriving at (non-Abelian) anomalies in $2n$ dimensions via the descent equations~\cite{Zumino:1983rz, Manes:1985df, Jackiw:1983nv, Zumino:1983ew, Stora:1983ct} (see also~\cite{Harvey:2005it, Weinberg:1996kr}). As an intermediate step of the descent formalism, one finds that the relevant quantity in $2n+1$ dimensions is the Chern-Simons (CS) action. We review all this in appendix~\ref{sec:review} and~\ref{sec:Cartan_homotopy_formula}, and discuss several important properties of anomaly polynomials, which we use frequently in later sections. 

A deeper physical insight of the descent procedure was then realized by the idea of \emph{anomaly inflow} originally discovered in~\cite{Callan:1984sa}. In its perturbative version, the bulk CS theory is not gauge-invariant when the theory is defined on a manifold with boundary. This variance is then cancelled by fermions localized on the boundary, making the overall bulk plus boundary theory consistent. A non-perturbative version of the anomaly inflow was discussed in~\cite{Witten:2015aba, Witten:2016cio} in the context of topological phases of matter, and its full justifications were presented recently in~\cite{Witten:2019bou} (see also~\cite{Yonekura:2016wuc}). Many condensed matter systems that exhibit non-trivial topological phases can be described by and hence are the physical realizations of anomaly inflow~\cite{Witten:2015aba}. One of the first example of topological phases of matter is the integer quantum Hall effect~\cite{Thouless:1982zz}, which admits a description in terms of $U(1)$ CS theory in the $(2+1)$D bulk of the material~\cite{Witten:2015aoa, Yonekura:2016wuc}:
\beq
S \sim \kappa \int A \wedge F, \;\; \kappa \in \mathbb{Z}.
\eeq
This is not gauge-invariant if the material has a physical boundary, and the resulting variance coincides with the first Chern class $\propto \int \frac{i}{2\pi} F$. This contribution is cancelled by the chiral anomaly of localized fermions on the $(1+1)$D boundary charged under $U(1)$ (called ``edge modes''). The combined system of bulk $+$ boundary is anomaly-free and the quantized Hall conductance is proportional to $\kappa$. 

In the language of anomaly inflow, 11D M-theory can also be thought of as being ``topological matter''. In the bulk of 11D spacetime, the 3-form superpartner of the metric field $C$ (with 4-form field strength $G = d C$) has the coupling
\beq
S \sim \int_{\rm 11D} C \wedge G \wedge G,
\eeq
and again this is not invariant in the presence of the boundary. The anomaly inflow is then completed by the introduction of ``edge modes'' which in this case are shown to be a 10D $E_8$ gauge supermultiplet~\cite{Horava:1995qa, Horava:1996ma}.

For completeness, we also mention that there have been many works with regard to anomalies in orbifold field theories, see for example~\cite{ArkaniHamed:2001is, Scrucca:2001eb, vonGersdorff:2003dt, Scrucca:2004jn, Barbieri:2002ic, Pilo:2002hu, Gripaios:2007tk}. Anomalies in orbifold gauge theories have a ``shape'' independent of extra-dimensional profiles and are localized on the boundaries. 
It may be instructive to recall the form of the anomaly in a theory with $S^1 / \left( \mathbb{Z}_2 \times \mathbb{Z}^\prime_2 \right)$ orbifold. Consider a $U(1)$ gauge theory with bulk fermions with boundary conditions (BC) denoted as $\left( \alpha_0, \alpha_1 \right)$, where $\alpha_i = \pm$ for $i=0(1)$ means Neumann ($+$) or Dirichlet ($-$) BC for the left-handed Weyl component of the 5D spinor at the UV (IR) boundary brane. For instance, $(+,+)$ gives rise to a left-handed zero mode, while the choice $(-,-)$ results in a right-handed zero-mode. Other choices do not lead to any zero mode. In this notation, the anomaly of a bulk fermion is given by~\cite{Gripaios:2007tk} 
\beq
\mathcal{A} (x,z) = \left( \frac{Q^3}{48 \pi^2} F_{\mu\nu} \tilde{F}^{\mu\nu} \right) \frac{1}{2} \left[ \alpha_0 \delta (z-z_0) + \alpha_1 \delta (z-z_1) \right].
\eeq
Here, the expression in parenthesis is the 4D $U(1)$ chiral anomaly with charge $Q$. The corresponding 5D anomaly is split between UV and IR boundaries with equal size, and the sign is determined by the BC.
From this, one sees that for $(+,+)$ or $(-,-)$ (like in the case of a $S^2/\mathbb{Z}_2$ orbifold), the anomaly function integrated over the extra dimension does not vanish. Such an anomaly in the orbifold theory is often referred to as a ``globally non-vanishing'' anomaly, and it arises from the zero mode. In this case, the cancellation of 4D anomalies is sufficient to render the full 5D theory consistent~\cite{ArkaniHamed:2001is}. For other choices of BC such as $(+,-)$ or $(-,+)$, however, the 5D anomaly function integrates to zero and is denoted as ``globally vanishing'' anomaly. Importantly, in this case, 4D anomaly cancellation is not enough, and in fact, there is no anomaly in the 4D effective field theory (EFT)~\cite{Scrucca:2001eb, Gripaios:2007tk}. While the 4D EFT is free of anomaly, the 5D theory nonetheless has localized anomalies. These anomalies may be cancelled by localized fermions, or localized higher-dimensional Wess-Zumino counter terms~\cite{Gripaios:2007tk}. Sometimes, they are cancelled by bulk Green-Schwarz mechanisms~\cite{vonGersdorff:2003dt, Scrucca:2004jn}. In some cases, anomaly inflow by the bulk CS action can restore consistency of the theory. In this sense, anomalies in orbifold gauge theories are certainly related to anomaly inflow, but the two are not equivalent.

Ever since its first discovery~\cite{Callan:1984sa}, interest in anomaly inflow was revived and has increased recently both in the condensed matter theory and high energy theory communities, partly thanks to the powerful machinery of the Dai-Freed theorem~\cite{Dai:1994kq} and generalized symmetries~\cite{Gaiotto:2014kfa}. See for instance~\cite{Witten:2015aba, Witten:2016cio} for the exposition of the subject written in a high energy theorist's language. Many surprising results are revealed by thorough explorations of anomaly inflow using modern techniques. Of course, CS theory in the context of AdS/CFT and its implications in terms of anomaly in the $SU(4)$ $R$-symmetry current were discussed in~\cite{Witten:1998qj}. Subsequently, the AdS/CFT interpretation of trace and chiral anomaly was studied for example in~\cite{Freedman:1998tz, Henningson:1998gx, Aharony:1999rz, Blau:1999vz, Nojiri:1999mh, Bilal:1999ph}. 

In the case of non-supersymmetric theories, AdS/CFT for brane-localized anomalies coming either from the bulk CS action or bulk fermions were studied for example in~\cite{Panico:2007qd, Gripaios:2007tk, Gripaios:2008ei}. In particular,~\cite{Gripaios:2007tk} discusses in detail a $U(1)$ theory with brane-localized scalars that cancel the anomaly from the bulk, and gives its CFT dual description. References~\cite{Gripaios:2007tk, Gripaios:2008ei} mention also the relevance for models of electroweak symmetry breaking and possible generalizations to non-Abelian bulk gauge groups. \cite{Panico:2007qd} studies in detail the case of holographic QCD-like theories and the low energy WZW action from the bulk CS action.

In this paper, we carefully analyze the physics of anomaly inflow in $\textnormal{AdS}_5$, and in particular obtain its holographic dual interpretations in terms of strongly coupled 4D (deformed) CFT. Our main contribution is, in addition to several new results reported below, to provide a unified and comprehensive discussion of the subject, together with a detailed description of the dual CFT picture for each case. To this end, we employ the gauge-fixed holographic partition function introduced in~\cite{Panico:2007qd} and generalize it to arbitrary UV and IR BCs. Moreover, we use the language of the Stora-Zumino descent equations~\cite{Zumino:1983rz, Manes:1985df, Jackiw:1983nv, Zumino:1983ew, Stora:1983ct}, which makes the treatment of non-Abelian groups particularly simple. 

While the anomaly inflow itself is largely insensitive to the specific form of the metric, the choice of $\textnormal{AdS}_5$ for the bulk metric nevertheless has a nice advantage. Namely, in this case, in addition to the standard features of anomaly inflow, i.e.~cancellation of ``problems'' between the bulk and boundary physics, a different description emerges. That is, the combination of bulk $+$ boundary can be interpreted as a holographic 4D conformal field theory. The dual 4D CFT consists of a strongly interacting CFT and external degrees of freedom, and the physics of anomaly inflow is dual to the interplay between these two sectors in a story of shared anomalies. 

 We found the introduction of UV and IR boundaries (branes) to lead to a remarkably rich description in the context of AdS/CFT. Namely, with Neumann BC, the UV brane allows to introduce the concept of ``weakly gauging'' of the global symmetry of the 4D CFT, resulting in ``dynamical'' source fields. On the other hand, the IR brane makes it possible to incorporate the phenomena of ``confinement'' and ``spontaneous symmetry breaking''. We also note that most of particle physics models in $\textnormal{AdS}_5$ were constructed with both UV and IR branes~\cite{Randall:1999ee} or even with additional branes~\cite{Agashe:2016rle}, and understanding the holographic version of anomaly inflow in a similar setup can provide more straightforward applications in particle physics. 

In fact, we have found that anomaly inflow with Neumann IR-BC in $\textnormal{AdS}_5$ is the holographic realization of `t Hooft anomaly matching. To this end, it was important to realize that in this case the variance generated by the bulk CS theory is cancelled by \emph{localized} degree of freedom on the boundaries. In the end, we show how anomaly matching occurs for both purely global (`t Hooft type) and weakly gauged (ABJ type) symmetries. These results, therefore, led us to speculate that fermion anomalies from holographic anomaly inflow are necessarily free of mixed anomalies with confining large-$N_s$ gauge symmetries of the dual CFT. These are discussed in section~\ref{sec:anomaly_inflow_wo_SSB}.  

On the other hand, when the IR-BC is chosen so that a subgroup $H_1$ of the bulk gauge group $G$ satisfies Neumann BC, while the coset $G/H_1$ is set to Dirichlet BC, the 4D dual theory describes spontaneous symmetry breaking $G \to H_1$ (whether weakly gauged or not). In this case, we show that in the low energy effective theory, the WZW action naturally appears and achieves anomaly matching. One key difference compared to the previous case with fully Neumann IR-BC is that in the current choice there are no IR brane-localized terms that are required by 5D consistency. This then is related to the fact that anomaly cancellation is done not by IR brane-\emph{localized} degrees of freedom, but rather by \emph{delocalized} modes, i.e.~Goldstone modes which holographically correspond to a 5D Wilson line $\Sigma \sim {\rm exp} \left( i \int d z A_z (x,z) \right)$. This may represent a less conventional feature of anomaly inflow. Discussion along these lines is presented in section~\ref{sec:SSB_WZW}.

To complete the outline of this paper and list additional new results, we first mention that in section~\ref{sec:gauge_theory_in_a slice_of_AdS5} we describe the basic setup of the gauge theory with CS action in a slice of $\textnormal{AdS}_5$. 
We then in section~\ref{sec:anomaly_inflow} describe the non-perturbative version of anomaly inflow in terms of the APS $\eta$-invariant~\cite{Atiyah:1975jf}, reducing it eventually to the perturbative version by means of the APS index theorem~\cite{Atiyah:1975jf}. 
In section~\ref{sec:holo_Z}, we perform the gauge fixing in a way that allows the holographic aspects of the theory to become apparent. In particular, we compute the holographic partition function, which we use in later sections to study anomaly inflow. 
In section~\ref{sec:quantization} we discuss the issue of quantization conditions for the CS level. In particular, we give there a holographic formulation of Witten's quantization condition for the WZW action. In addition, we argue that such a condition may be achieved from a ``deformed'' theory related to the original theory by changes of BCs in 5D. In several appendices, after listing our notation and conventions, we provide a detailed review of chiral anomalies described in terms of descent equations. We also discuss Cartan's homotopy formula and show our derivation of the shifted CS action and associated shifted anomaly and local counter terms. We then list important properties of the shifted CS action and local counter terms, which we make frequent use of in the study of anomaly inflow. Furthermore, we discuss the WZW action in terms of the shifted CS action. Our hope is that the inclusion of this rather extensive review (with some new derivations) makes the presentation self-contained and useful.

\section{Gauge theory in a slice of $\textnormal{AdS}_5$}
\label{sec:gauge_theory_in_a slice_of_AdS5}

In the rest of the paper, we wish to study a gauge theory in a slice of five-dimensional anti-de Sitter space, $\textnormal{AdS}_5$.
In particular, our prime interest is to study a gauge theory in the presence of a Chern-Simons action and to understand its 4D CFT dual interpretation. 
Most of what we discuss will go through whether or not the theory includes the usual gauge kinetic terms.
In the absence of the gauge kinetic term, the 5D theory is purely topological. 
As we will discuss in section~\ref{sec:anomaly_inflow} such a topological field theory may be thought of as being induced by integrating out massive Dirac fermions in the bulk \emph{perturbatively}~\cite{Redlich:1983kn, Redlich:1983dv, Dunne:1998qy}. We, however, do not need to specify a particular UV completion. 

When the theory includes both the gauge kinetic and CS terms, the action for the gauge group $G$ is given by
\beq
\begin{gathered}
S = S_0 + S_{\rm CS}, \qquad \qquad%
S_0 = - \frac{1}{2g_5^2} \int d^5x \sqrt{g} \Tr[ F_{MN} F_{LS} g^{ML} g^{NS}], \\
S_{\rm CS} = c \int d^5x \; \epsilon^{MNPQR} \Tr( A_M \partial_N A_P \partial_Q A_R + \frac{3}{2} A_M A_N A_P \partial_Q A_R + \frac{3}{5} A_M A_N A_P A_Q A_R), 
\end{gathered}
\eeq
where the metric takes the form (with mostly minus convention, $\eta_{\mu\nu} = \diag(+,-,-,-)$)
\beq
ds^2 = a(z)^2 \left( d x^2 - dz^2 \right), \qquad a(z) = \frac{1}{k z}.
\eeq
Here, $k$ is the compactification scale. The theory is defined on the interval $z_0 \leq z \leq z_1$. At the boundaries, we put 3-branes that we call \emph{UV brane} and \emph{IR brane}, respectively.

It is often convenient to adopt a differential form notation to simplify expressions. In these terms, the gauge connection is a 1-form $A \equiv A_M^A T^A dX^M$ and the field strength is a 2-form $F = d A + A^2 = \frac{1}{2} F_{MN}^A T^A dX^M dX^N$. In this notation, the wedge product symbols are suppressed, e.g.~$A^2 = A \wedge A$. For instance, the CS action can be written as
\beq
S_{\rm CS} = c \int  \Tr( A dA dA + \frac{3}{2} A^3 dA + \frac{3}{5} A^5).
\eeq
At the moment, the overall coefficient $c$, called the \emph{level}, is considered to be a free parameter. In section~\ref{sec:quantization}, we will show that it is quantized by anomaly cancellation requirements.
As reviewed in appendix~\ref{subsec:descent_eq}, the 5D CS action is the integral of the \emph{canonical} form $\omega_5^{(0)} (A)$,
\beq
S_{\rm CS} = c \int_{\rm 5D} \omega_5^{(0)} (A). 
\label{eq:S_CS_in_terms_of_omega_5}
\eeq
The canonical CS term $\omega_5^{(0)} (A)$ is related to the 6D Abelian anomaly $\Omega_6 (A) = {\rm Tr} \left(F^3 \right)$ and 4D (non-Abelian) chiral anomaly $\omega_4^{(1)} (\alpha, A)$ via descent equations as
$\Omega_6 = d \omega_5^{(0)} (A)$ and $\delta_\alpha \omega_5^{(0)} (A) = d \omega_4^{(1)} (\alpha, A)$.
One has the freedom to add local counter terms to this, called $d B_4$ in appendix~\ref{subsec:descent_eq} and~\ref{sec:Cartan_homotopy_formula}, and this leads to the \emph{shifted} CS action, $\omega_5^{(0)} \to \tilde{\omega}_5^{(0)} = \omega_5^{(0)} + d B_4$. The resulting 4D anomaly is then a shifted anomaly.
 
Suitable gauge-invariant boundary conditions (BC) can be obtained by studying the variation of $S_0$. This leads to 
\begin{align}
\delta S_0 &= \frac{2}{g_5^2} \int dz d^4x \; \Tr[ \delta A^\mu \left( a D^\nu F_{\nu\mu} + D_z (a F_{\mu z}) \right)] + \Tr[ \delta A^z D^\mu \left( a F_{\mu z} \right)] \label{eq:bulk_EOM} \\
&\phantom{{}={}} + \frac{2}{g_5^2} \int_{\rm UV} d^4 x \; \Tr[ \delta A^\mu a F_{\mu z}] - \frac{2}{g_5^2} \int_{\rm IR} d^4 x \; \Tr[ \delta A^\mu a F_{\mu z}]. \label{eq:BC}
\end{align}
Eq.~(\ref{eq:bulk_EOM}) gives the bulk equations of motion (EOM), while eq.~(\ref{eq:BC}) is the boundary EOM and provides the BC. A vanishing boundary variation is achieved by either setting $\delta A^\mu = 0$ (Dirichlet BC) or $F_{\mu z} = 0$ (Neumann BC).\footnote{In the axial gauge $A_z=0$, the condition $F_{\mu z} = 0$ becomes $\partial_z A_\mu = 0$, a usual form of Neumann BC. Also, once a BC for $A_\mu$ is specified, that of $A_z$ is fixed by the bulk EOM.} While in the orbifold construction $S^1/\mathbb{Z}_2$ the UV and IR BC are correlated, we are treating the extra-dimensional direction as an interval (or $S^1/ \left( \mathbb{Z}_2 \times \mathbb{Z}^\prime_2 \right)$) and we are free to choose UV and IR BC independently.

\section{Anomaly inflow}   
\label{sec:anomaly_inflow}

In this section, we review the basic idea of anomaly inflow. 
However, we first remark that the discussion in the rest of paper can be understood in a mostly self-contained manner with a minimal conceptual picture of anomaly inflow. Readers primarily interested in anomaly inflow by bulk CS theory and its AdS/CFT interpretations, therefore, may read a brief discussion below around eq.~(\ref{eq:U(1)_anomaly_inflow_1}) and (\ref{eq:U(1)_anomaly_inflow_2}) and safely skip the rest of this section. 

A non-perturbative formulation of anomaly inflow was given in a recent paper~\cite{Witten:2019bou}.
In the non-perturbative version, anomaly inflow is described in terms of the $\eta$-invariant of Atiyah, Patodi, and Singer (APS)~\cite{Atiyah:1975jf}. The idea is the following: we start with a bulk \emph{massive} fermion $\Psi (x,z)$ coupled to a background gauge field and gravity. We, however, will focus only on the gauge field part. The goal is to compute the partition function of this theory with some \emph{local} (and chiral) BC. Examples of such local BC are $P_{\rm R} \Psi |_{\rm UV} = 0$ and $P_{\rm L} \Psi |_{\rm IR} = 0$ and so on, where $P_{\rm L/R} = \frac{1}{2} \left( 1 \mp \gamma^5 \right)$ is the chiral projection operator. Since we have two boundaries, and adopting the interval formulation (or $S^1 / \left( \mathbb{Z}_2 \times \mathbb{Z}^\prime_2 \right)$), we impose independent UV and IR BC. We denote the BC that keeps the LH component of $\Psi$ (i.e.~RH component projected out) as ``$L$'', while the BC that keeps the RH component as ``$R$''. For instance, writing a general BC as $\left( \alpha_0, \alpha_1 \right)$, the choice $\left( L, R \right)$ means that the bulk $\Psi$ is projected onto LH (RH) component on the UV (IR) brane. Calling the bulk $Y$ and boundaries $W_0$ and $W_1$, the path integral of a bulk massive fermion with the BC $\left( \alpha_0, \alpha_1 \right)$ may be written as\footnote{See~\cite{Yonekura:2016wuc, Witten:2019bou} for the description of path integral in terms of state overlaps and for more details.}
\beq
Z [Y, \left( \alpha_0, \alpha_1 \right)] = \langle \alpha_0 \vert Y \vert \alpha_1 \rangle \equiv \int_{\left( \alpha_0, \alpha_1 \right)} \; \mathcal{D} \Psi \mathcal{D} \bar{\Psi} \; e^{-S}.
\label{eq:Z_Y}
\eeq
The states $\vert \alpha_{0 (1) } \rangle$ defined on the Hilbert space of the boundary incorporate the BC. The meaning of this statement is as follows. Let us consider the example of $L$-UV-BC. Writing the boundary Dirac operator as $\mathcal{D}_W$ and chirality operator as $\gamma^\tau$ with $\{ \gamma^\tau, \mathcal{D}_W \} = 0$, the bulk fermion $\Psi$ can be expanded in terms of modes of $\mathcal{D}_W$,
\beq
\mathcal{D}_W \psi_{L,a} = \lambda_a \psi_{R,a}, \; \mathcal{D}_W \psi_{R,a} = \lambda_a \psi_{L,a}, \; \gamma^\tau \psi_{L/R,a} = \mp \psi_{L/R,a} .
\eeq
The mode expansion is given by
\beq
\Psi = \sum_a \left( A_{L,a} \psi_{L,a} + A_{R,a} \psi_{R,a} \right),
\eeq
where the fermionic operators coefficients $A_{L/R,a}$ are annihilation operators for LH (RH) modes. 
In terms of this, $L$-UV-BC is a condition on the UV value such that the RH modes are fixed to be zero, while the LH mode is unconstrained. The latter condition may be rephrased as its ``conjugate momentum'' mode being fixed to be zero.\footnote{Recall that a plane wave in position space corresponds to a delta function in its conjugate momentum space, and vice versa.} Hence, $\langle \alpha_0 = L \vert$ is defined by the properties
\beq
\langle L \vert A_{R,a} = 0, \;\; {\rm and} \;\; \langle L \vert A_{L,a}^\dagger = 0.
\eeq

Going back to the evaluation of the partition function eq.~(\ref{eq:Z_Y}), we note that, in general, the bulk Dirac operator is not self-adjoint in the presence of a boundary. In order for the Dirac operator to be self-adjoint, hence to result in a well-defined real spectrum, a special class of \emph{global} BC, called APS-BC~\cite{Atiyah:1975jf}, should be imposed. Without going into too much technical details, we simply quote the result here, referring to~\cite{Yonekura:2016wuc, Witten:2019bou} for more details. Provided the geometry near the boundary is $\sim [z_0, \epsilon ) \times W_0$ (similarly for the IR boundary) and $\vert m \vert \epsilon \gg 1$ (i.e.~the bulk mass $m$ is much larger than the curvature scale), the Euclidean path integral near the boundary is effectively a projection to the vacuum $e^{-\epsilon \vert m \vert} \approx \vert \Omega \rangle \langle \Omega \vert$ and the partition function eq.~(\ref{eq:Z_Y}) can be computed in the limit $\vert m \vert \to \infty$ to get
\beq
\begin{aligned}
Z [Y, \left( \alpha_0, \alpha_1 \right)] &= \langle \alpha_0 \vert \Omega \rangle \langle \Omega \vert Y \vert \Omega \rangle \langle \Omega \vert \alpha_1 \rangle \\ 
&= \frac{\langle \alpha_0 \vert \Omega \rangle \langle \Omega \vert {\rm APS} \rangle}{\vert \langle \Omega \vert {\rm APS} \rangle \vert^2} \; \langle {\rm APS} \vert Y \vert {\rm APS} \rangle \; \frac{\langle {\rm APS} \vert \Omega \rangle \langle \Omega \vert \alpha_1 \rangle}{\vert \langle \Omega \vert {\rm APS} \rangle \vert^2}. 
\label{eq:Z_Y_UV_bulk_IR}
\end{aligned}
\eeq
We used $e^{-\epsilon \vert m \vert} \approx \vert \Omega \rangle \langle \Omega \vert$ and that in the limit $\vert m \vert \to \infty$ the Hilbert space becomes effectively one-dimensional. In particular, this implies $\vert Y \rangle \propto \vert {\rm APS} \rangle \propto \vert \Omega \rangle$, where $\vert Y \rangle$ is the path integral over $Y$ (recall that path integration over $Y$ gives a state vector in the Hilbert space on the boundary). The first and the last factor in eq.~(\ref{eq:Z_Y_UV_bulk_IR}) are UV and IR boundary contributions and the middle term is identified as the bulk contribution. $\vert {\rm APS} \rangle$ is a state in the Hilbert space of the boundary that incorporates the required APS-BC. As shown in~\cite{Witten:2019bou}, the boundary contributions are related to $\left\vert {\rm Det} \mathcal{D}_W^{\pm} \right\vert$, where $\mathcal{D}_W^{\pm}$ is the chiral Dirac operator on the boundary $W_{0(1)}$ and its exact nature depends on the BC and associated localized modes.\footnote{A study of non-perturbative anomaly inflow in extra-dimensional particle physics models and its connection with ``anomalies in orbifold field theories'' will be presented elsewhere~\cite{SH}.} For us, the precise form of the boundary contributions are not important.\footnote{To be clear, the precise form of the boundary term is very important. In fact, because the boundary term is $\left\vert {\rm Det} \mathcal{D}_W^{\pm} \right\vert$, and not say $ {\rm Det} \mathcal{D}_W^{\pm} $, the non-perturbative anomaly inflow formula eq.~(\ref{eq:non-pert_anomaly_inflow_formula}) provides the unique determination of the phase of the boundary partition function, provided eq.~(\ref{eq:non-pert_anomaly_inflow_formula}) is independent of the choice of $Y$ (i.e.~absence of anomaly).  } Instead, we are interested in the bulk contribution, $\langle {\rm APS} \vert Y \vert {\rm APS} \rangle$. This is just a path integral over $Y$ with APS-BC. As mentioned above, the bulk Dirac operator $\mathcal{D}_Y$ with APS-BC is self-adjoint and its spectrum is well-defined and real. The massive fermion now can be integrated out non-perturbatively, and the phase of the partition function can be determined. In fact, this is one of the important reasons to rewrite the partition function in the form of eq.~(\ref{eq:Z_Y_UV_bulk_IR}). Formally, the path integral of $\Psi$ with negative mass $-m \; ( m>0 )$ is ${\rm Det} \left( \mathcal{D}_Y - i m \right)$. Using Pauli-Villars (PV) regularization (with positive mass $+M \; ( M>0 )$), we get
\beq
\langle {\rm APS} \vert Y \vert {\rm APS} \rangle = \prod_a \frac{\lambda_a - im }{\lambda_a + i M} = \prod_a \frac{\lambda_a}{\lambda_a + i M} \; \prod_a \frac{\lambda_a - i m}{\lambda_a}.
\label{eq:partition_function_massive_fermion}
\eeq
Each factor of the form $\prod_a \lambda_a / (\lambda + i M)$ is recognized as the partition function of a massless fermion in PV-regularization. In the limit $M \gg \vert \lambda_a \vert$, each term has a phase $- i \frac{\pi}{2} {\rm sign} (\lambda_a)$ and the overall phase is therefore $ - i \frac{\pi}{2} \sum_a {\rm sign} (\lambda_a)$. This formal expression, which requires a regularization, is known as the APS $\eta$-invariant. Similarly, in the limit $m \to \infty$, the phase of the second factor is found to be also  $ - i \frac{\pi}{2} \sum_a {\rm sign} (\lambda_a)$. Therefore, the phase of the bulk term is given by\footnote{The phase of eq.~(\ref{eq:partition_function_massive_fermion}) is indeed regularized. For $\lambda_a \gg M, m$, the phase is 0 and the sum over the entire ``UV'' modes is trivial. In addition, for the  zero mode (if the Dirac operator does have zero modes), the phase is given by (setting $m = M \to \infty$) $-i \pi$. In this sense, in PV regularization, ${\rm sign} (\lambda_a = 0) = +1$.}
\beq
{\rm phase \; of \;} \langle {\rm APS} \vert Y \vert {\rm APS} \rangle = - i \pi \sum_a {\rm sign} (\lambda_a) \vert_{\rm reg} = - i \pi \eta_Y.
\label{eq:phase_eta}
\eeq
If we instead had chosen a positive mass $+m$, the phase of the PV-regularized (with negative mass $-M$) partition function would be just minus that of eq.~(\ref{eq:phase_eta}). Using this, the integration of the bulk massive fermion non-perturbatively eq.~(\ref{eq:Z_Y_UV_bulk_IR}) may be written as 
\beq
Z [Y, \left( \alpha_0, \alpha_1 \right)] = Z_0 \cdot e^{\mp i \pi \eta_Y} \cdot Z_1,
\label{eq:non-pert_anomaly_inflow_formula}
\eeq
where $Z_{0 (1)}$ are the boundary contributions, while the exponentiated $\eta$-invariant is the bulk contribution. 
This is the formula for the non-perturbative version of anomaly inflow. A perturbative version can be obtained from this thanks to the APS index theorem~\cite{Atiyah:1975jf}. For a closed manifold $X$ of dimension $d+2$\footnote{We imagine a theory defined on a $d$-dimensional spacetime $W$ ($d=4$ in our case), which itself is a boundary of a $d+1$ manifold $Y$ (bulk of $\textnormal{AdS}_5$). We then define a $d+2$-dimensional manifold $X$, whose boundary is $\partial X = Y \cup - Y'$; since $Y$ has a boundary, we combine $Y$ with $Y'$ where $\partial Y' = W = \partial Y$ to form a closed manifold $Y \cup - Y'$ (where $- Y'$ is the orientation reversal of $Y$).  }, according to the Atiyah-Singer index theorem~\cite{Atiyah:1968mp}, the index of the Dirac operator, the number of positive chiral zero modes minus the number of negative chiral zero modes, is equal to
\beq
{\rm ind} \left( \mathcal{D}_X \right) = n_+ - n_- = \int_X I_{d+2}, \;\;\;\; I_{d+2} = \left. \hat{A} (R) {\rm tr \; exp} \left( \frac{i F}{2\pi} \right) \right\vert_{d+2},
\eeq
where $\hat{A} (R)$ is the Dirac genus of the manifold ($R$ being the Ricci curvature)\footnote{For an explicit form of $\hat{A} (R)$ see e.g.~\cite{AlvarezGaume:1984dr, AlvarezGaume:1984nf}.} and ${\rm tr \; exp} \left( \frac{i F}{2\pi} \right)$ is the Chern character ($F$ is the field strength 2-form). If, on the other hand, the manifold $X$ has a boundary, the relevant formula is the APS index theorem~\cite{Atiyah:1975jf},
\beq
{\rm ind} \left( \mathcal{D}_X \right) = \int_X I_{d+2} - \frac{\eta_{\partial X}}{2},
\eeq
where the $\eta$-invariant contribution may be interpreted as the boundary correction. In the perturbation theory we are focusing on here, it is straightforward to see that (taking $d=4$)
\beq
e^{i \pi \eta_Y} = {\rm exp} \left( i \frac{1}{48 \pi^2} \int_Y \omega_5^{(0)} (A) \right),
\eeq
that is, the bulk term in eq.~(\ref{eq:non-pert_anomaly_inflow_formula}) turns into the 5D CS action in the perturbative limit. The perturbative anomaly inflow is performed by the interplay of the bulk CS interaction and the boundary modes. In fact, this is our starting point for the rest of the paper. We study the anomaly inflow of various CS theories and obtain its dual CFT descriptions.

In order to illustrate the point, let us consider the simple example of a $U(1)$ CS theory in a slice of $\textnormal{AdS}_5$:
\beq
S_{\rm CS} = c \int_{\rm 5D} \; A dA dA = c \int_{\rm 5D} \; A F F.
\label{eq:U(1)_anomaly_inflow_1}
\eeq
%
Under a gauge transformation, $A \to A + d v$ and $F \to F$, and CS action changes by
\beq
\delta S_{\rm CS} = c \int_{\rm IR} \; v F F - c \int_{\rm UV} \; v F F.
\label{eq:U(1)_anomaly_inflow_2}
\eeq
To get eq.~(\ref{eq:U(1)_anomaly_inflow_2}) we performed an integration by parts and used $d F =0$. We see that the CS theory defined on a manifold with boundary is not gauge invariant, and the variance terms induced on the boundary have the form of chiral anomaly. In order to make the 5D theory consistent, such surface terms need to be cancelled, and one way to achieve this is to introduce boundary localized fermions (edge modes) charged under $U(1)$. 
This is the essential feature of perturbative anomaly inflow and can be generalized to more complicated (and interesting) cases. Below, we study various CS theories, including non-Abelian and/or mixed CS theories, with special emphasis on their holographic dual descriptions. While the description given so far makes the interplay between bulk and boundary physics clear, it is not yet the optimal one for a holographic study. 
For this reason, in the next section, we first discuss a description of the gauge theory in which the AdS/CFT duality becomes apparent.

\section{Holographic partition function}   
\label{sec:holo_Z}

In this section, we discuss the gauge fixing issue, adopting the axial gauge. Since we are interested in the holographic study, we perform the gauge fixing in such a way that the holographic nature of the $\textnormal{AdS}_5$ gauge theory becomes manifest: we derive the gauge-fixed holographic partition function $Z^{\rm g.f.}$. Our discussion in this section is influenced by the work of~\cite{Panico:2007qd}. We, however, generalize it to arbitrary UV and IR boundary conditions. In addition, when the UV-BC is chosen to be Neumann for some non-trivial subgroup $H_0 \subset G$, we show that there is a residual gauge redundancy, requiring further (brane-localized) gauge fixing.\footnote{In~\cite{Hirn:2005nr}, in the context of AdS/QCD without a CS term, a discussion was presented in which a field redefinition, instead of gauge fixing, is used. Both approaches produce agreeing results when expected.}

Let us consider a very general situation.
We take the bulk gauge group to be $G$. We imagine a general BC by which $G$ is broken down to $H_0 \subset G$ on the UV brane and to $H_1 \subset G$ on the IR brane. In this case, the full 5D gauge symmetry is given by
\beq
G_B = \{ g(x,z) \in G \; \vert \; \hat{g} \equiv g(x,z_0) \in H_0, \bar{g} \equiv g(x,z_1) \in H_1 \}.
\label{eq:G_B}
\eeq
The dual 4D CFT then has a global symmetry group $G$, which is spontaneously broken to $H_1$ by confinement at the scale associated with the IR brane location. $H_0 \subset G$ on the UV brane is dual to the fact that the $H_0$ part of $G$ is weakly gauged, featuring an explicit breaking of $G$ by gauging. The inclusion of the 5D CS action further incorporates the anomaly structure into the 4D dual gauge theory. 

We denote the Lie algebra of $G$ as $\mathbf{g}$. Similarly, $\mathbf{h}_0$ is the Lie algebra of $H_0$ and we denote the space generated by the coset $G/H_0$ generators to be $\mathbf{k}_0$. Likewise, $\mathbf{h}_1$ is the Lie algebra of $H_1$ and $\mathbf{k}_1$ denotes the space spanned by the generators of the coset $G/H_1$. Generators of $\mathbf{g}$ are written as $T^A \in \mathbf{g}, A=1, \dots, {\rm Dim}[\mathbf{g}]$. Likewise, $T_m^i \in \mathbf{h}_m$ and $ T_m^a \in \mathbf{k}_m, m=0,1$ are the unbroken and broken generators, respectively.

Before gauge fixing, the holographic partition function of a gauge theory of eq.~(\ref{eq:G_B}) takes the form
\begin{align}
& Z \left[B^a\right] = \int \mathcal{D} B^i \; \mathcal{Z} \!\left[B^A\right], \label{eq:Z_h_gauge_unfixed} \\
& \mathcal{Z} \left[B^A\right] = \int \mathcal{D} A_\mu (x,z) |^{\hat{A} = B}_{\bar{A} :  \left[
\begin{array}{l}
\scriptscriptstyle (F)_1^i = 0 \\
\scriptscriptstyle (A)_1^a = 0
\end{array} \right. }
\mathcal{D} A_z (x,z) \; e^{i S[A_\mu (x,z), A_z (x,z)]}.
\label{eq:cal_Z_B_A}
\end{align}
Let us explain the notation we used in these expressions. First, $\mathcal{Z} \left[B^A\right]$ is the partition function with UV boundary value of the bulk gauge field $A$ taken to be $B^A$. Here, we suppressed the Lorentz index for the sake of brevity. The superscript $A$ runs over \emph{all} generators. This statement about the UV-BC is also written schematically as the superscript ``$\hat{A} = B$'' in eq.~(\ref{eq:cal_Z_B_A}). The subscript in the same equation denotes instead the IR-BC. $(F)_1^i = 0$ means $F_{\mu z} = 0$ for $T_1^i \in \mathbf{h}_1$. 
The expression $(A)_1^a=0$ can be understood in the same way.
As it is, $\mathcal{Z} \left[B^A\right]$ is the holographic partition function with Dirichlet UV-BC for all generators. 
In order to obtain the partition function corresponding to eq.~(\ref{eq:G_B}), we need to promote the background source for the $T_0^i \in \mathbf{h}_0$ to dynamical fields. This is done by path integrating over the fields $B^i$. Once this is done, then the partition function depends only on $B^a$, the background fields associated with $T_0^a \in \mathbf{k}_0$. The final result of this whole procedure is summarized by eq.~(\ref{eq:Z_h_gauge_unfixed}) and (\ref{eq:cal_Z_B_A}).

As a next step, we want to incorporate the axial gauge, $A_z=0$. As usual, this is done by inserting the following gauge-fixing factor into eq.~(\ref{eq:Z_h_gauge_unfixed}):
\beq
\begin{aligned}
1 &= \int \mathcal{D} A_z \; \delta (A_z) 
= \int \mathcal{D} g \; {\rm Det} \left[ \frac{\delta A^g_z}{\delta g} \right] \; \delta \left( A^g_z \right) \\
&= \int \mathcal{D} \Sigma_1 (x) \; \mathcal{D} h_1 (x) \; \mathcal{D} h_0 (x) \; \mathcal{D} g \vert^{\hat{g}=h_0}_{\bar{g}=\Sigma_1 \circ h_1} \; {\rm Det} \left[ D_z (A_z) \right] \; \delta \left( A^g_z \right),  
\label{eq:1_gauge_fixing}
\end{aligned}
\eeq
where $A^g \equiv g \left( d + A \right) g^{-1}$ is the gauge transformation of the gauge connection $A$ by $g \in G$. In here and in the following, we adopt the conventions of appendix~\ref{appendix:notation_conventions}. 
To obtain the last line, we used the fact that for a compact group $G$ and a closed subgroup $H$ the integration over the group manifold can be split into $H$ and $G/H$ parts with proper left invariant Haar measures~\cite{Preskill:1990fr, Panico:2007qd}. 
In particular, for the coset part, the $G$-invariant measure on $G/H$ can be expressed as~\cite{Preskill:1990fr}
\beq
\mathcal{D} \Sigma = \prod_a \left( \Sigma^{-1} d \Sigma \right)^a,
\eeq
where $\left( \Sigma^{-1} d \Sigma \right)^a$ is defined via $\left( \Sigma^{-1} d \Sigma \right)_k = \left( \Sigma^{-1} d \Sigma \right)^a T^a$, $T^a \in \mathbf{k}$. The $G$-invariance is seen by noting that under an arbitrary $g \in G$, $\left( \Sigma^{-1} d \Sigma \right)_k$ transforms as\footnote{On the other hand, $\left( \Sigma^{-1} d \Sigma \right)_h$ transforms inhomogeneously as
\beq
\left( \Sigma^{-1} d \Sigma \right)_h \to h^{-1} (g, \Sigma) \left( \Sigma^{-1} d \Sigma \right)_h h(g, \Sigma) + h^{-1} d h.
\eeq}
\beq
\left( \Sigma^{-1} d \Sigma \right)_k \to h^{-1} (g, \Sigma) \left( \Sigma^{-1} d \Sigma \right)_k h (g, \Sigma).
\eeq
In the above, $h (g, \Sigma) \in H$ is defined by $g \Sigma = \Sigma^g (g, \Sigma) h (g, \Sigma)$.
If we had used a naive (non-invariant) measure $\prod_a d \xi_a$ (recall $\Sigma = e^{-\xi_a T^a}$), then $G$-invariance of the quantum theory could be restored by adding a proper term $\int \mathcal{L} (\xi)$ to the action. The role of this term is to precisely cancel the non-invariance of the naive measure.

The gauge redundancy by $G_B$ can be singled out by the following change of variable: $g \to g = \tilde{\Lambda} \circ g'$, where $\hat{\tilde{\Lambda}} = \tilde{\Lambda} (x, z_0) = 1$ and $\bar{\tilde{\Lambda}} = \tilde{\Lambda} (x, z_1) = \Sigma_1 (x)$.\footnote{The existence of such $\tilde{\Lambda} (x,z)$ is guaranteed provided $\pi_4 (G/H) = 0$. This is understood by observing that $\tilde{\Lambda}$ is effectively an extension of $\Sigma_1 (x)$ defined on $S^4$ (IR brane) to a 5D disk $D_5$. If $\pi_4 (x)=0$, then every coset element admits an extension to $D_5$. If, on the other hand, $\pi_4 (G/H)$ is non-trivial, then while a $\Sigma (x)$ belonging to the trivial class of $\pi_4 (G/H)$ can be extended to $D_5$, any $\Sigma (x)$ in the non-trivial elements can at best be deformed into a representative of the corresponding class and the resulting object is defined on $S^4 \times [0,1]$~\cite{DHoker:1994rdl}. For simplicity, we assume $\pi_4 (G/H) = 0$ in what follows.} After inserting the transformed version of eq.~(\ref{eq:1_gauge_fixing}) into eq.~(\ref{eq:Z_h_gauge_unfixed}) we get
\begin{align}
& Z [B^a] = \int \mathcal{D} B^i \; \mathcal{Z} \left[B^A\right] \label{eq:Z_h_gauge_unfixed_2}, \\
& \mathcal{Z} \left[B^A\right] = \int \left[ \mathcal{D} h_0 \; \mathcal{D} h_1 \; \mathcal{D} g' \vert^{\hat{g'}=h_0}_{\bar{g'}=h_1} \right] \; \mathcal{D} \Sigma_1 (x) \mathcal{D} A_\mu (x,z) |^{\hat{A} = B}_{\bar{A} :  \left[
\begin{array}{l}
\scriptscriptstyle (F)_1^i = 0 \\
\scriptscriptstyle (A)_1^a = 0
\end{array} \right. }
\mathcal{D} A_z (x,z)  \nonumber \\
&\phantom{{}\mathcal{Z} \left[B^A\right]={}} \times \; {\rm Det} \left[ D_z (A_z) \right] \; \delta \!\left( A^{\tilde{\Lambda} \circ g'}_z \right) \; e^{i S[A_\mu (x,z), A_z (x,z)]}.
\label{eq:cal_Z_B_A_2}
\end{align}
Notice that $\left[ \mathcal{D} h_0 \; \mathcal{D} h_1 \; \mathcal{D} g' \vert^{\hat{g'}=h_0}_{\bar{g'}=h_1} \right]$ is nothing but the integration over $G_B$, the gauge redundancy we hope to remove from the path integral.

We proceed further with a change of variable: $A' = A^{\tilde{\Lambda} \circ g'}$. This simplifies the argument of the delta function, making the evaluation of the $A_z$-integral trivial. Taking into account the changes of UV-BC, IR-BC, and performing the $A_z$-integral using the delta function, we arrive at
\begin{align}
& Z [B^a] = \int_{G_B} \left[ \mathcal{D} h_0 \; \mathcal{D} h_1 \; \mathcal{D} g' \vert^{\hat{g'}=h_0}_{\bar{g'}=h_1} \right] \; \int \mathcal{D} B^i \; \mathcal{D} \Sigma_1 (x) \mathcal{D} A'_\mu (x,z) |^{\hat{A'} = B^{h_0}}_{\bar{A'} :  \left[
\begin{array}{l}
\scriptscriptstyle \left( (F'_1)^{\Sigma_1^{-1}} \right)^i = 0 \\
\scriptscriptstyle \left( (A'_1)^{\Sigma_1^{-1}} \right)^a = 0
\end{array} \right. }  \nonumber \\
&\phantom{{}\mathcal{Z} \left[B^A\right]={}} \times \;  e^{i S\left[(A'_\mu)^{(\tilde{\Lambda} \circ g')^{-1}} (x,z), 0^{(\tilde{\Lambda} \circ g')^{-1}}\right]}.
\label{eq:Z_h_gauge_unfixed_3}
\end{align}
To get eq.~(\ref{eq:Z_h_gauge_unfixed_3})  we used the fact that, upon $A_z$-integration, ${\rm Det} \left[ D_z (A_z) \right] \to {\rm Det} [\partial_z]$ and dropped this irrelevant constant. We also used the $H_1$-invariance of the IR-BC to simplify its form. Notice that now the second argument (fifth component) in the action is a pure gauge contribution, which in general does not vanish.

Let us now analyze how the latest form of the partition function depends on $h_0$ and $g'$. First of all, when $H_0$ is not trivial, there is a residual gauge freedom that needs to be fixed. The existence of this residual gauge redundancy is seen by the appearance of the extra integration over $h_0$ compared to the case with a trivial $H_0$. A more careful statement can be made as follows. The bulk axial gauge fixing coincides with the condition $A_z' = g ( \partial_z + A_z ) g^{-1} = 0$. This is a condition on $g$, selecting a specific gauge orbit. The solution to this equation is the Wilson line
\beq
g = P \; {\rm exp} \left( \int_{z_0}^z dz' \; A_z \left(x,z'\right) \right),
\eeq 
stretched from the UV brane to a point in the bulk at $z$.\footnote{An equivalent solution is the Wilson line from the IR brane to a bulk point at $z$. Our argument can be applied to both cases.} This $g$ successfully removes any $A_z$ component everywhere in the bulk except at $z=z_0$ where $g$ becomes 1. For the Dirichlet UV-BC, this is not an issue since the UV brane preserves no gauge symmetry. If, however, the UV-BC involves a non-trivial $H_0$, this indicates that the bulk axial gauge fixing is not complete, and we need to add a brane-localized gauge fixing term to eliminate the residual gauge freedom. Since the details of this extra gauge fixing do not affect our discussion below in any crucial way, we simply set $h_0=1$ and drop the integration over $h_0$. This is also equivalent to properly reinterpreting the gauge fixing constraint in eq.~(\ref{eq:1_gauge_fixing}), so that instead of integrating over the entire gauge manifold, we pick a specific gauge orbit.

The advantage of this approach is that it also allows us to more easily study the $g'$ transformation appearing in the action. We know that there are different contributions to $S[A_\mu,A_5]$, namely the gauge and CS action. The former is by assumption gauge invariant. The latter in general is not, but nevertheless we can use the fact that its variation only contributes as a boundary term to the variation of the full action, and therefore only depends on the boundary values of $g'$. Thus, by setting $h_0=1$, and by assuming from now on that
the CS action satisfies $\omega_5^{(0)} (A_h) = 0$, i.e.~the CS action (hence the associated anomaly $\omega_4^{(1)}$) vanishes when the gauge field $A$ is restricted to its $H_1$ part, we can conclude that the full action $S[A_\mu,A_5]$ is invariant under any $g'\in G_B$ transformation. The property $\omega_5^{(0)} (A_h) = 0$ is referred to as an anomaly-free embedding (AFE) of $H_1 \subset G$.

After dropping the $g'$ transformation from eq.~(\ref{eq:Z_h_gauge_unfixed_3}), we make one last change of variable, which moves the $\Sigma_1$ dependence from the IR-BC to the UV-BC. We set $A_\mu = (A_\mu')^{\Sigma_1^{-1}}$ to obtain
\begin{equation}
Z [B^a] = \int_{G_B} \left[ \mathcal{D} h_0 \; \mathcal{D} h_1 \; \mathcal{D} g' \vert^{\hat{g'}=h_0}_{\bar{g'}=h_1} \right] \; \int \mathcal{D} B^i \; \mathcal{D} \Sigma_1 (x) \mathcal{D} A_\mu (x,z) |^{\hat{A} = B^{\Sigma_1^{-1}}}_{\bar{A} :  \left[
\begin{array}{l}
\scriptscriptstyle  F_1^i = 0 \\
\scriptscriptstyle  A_1^a = 0
\end{array} \right. } 
\;  e^{i S\left[A^{\Lambda} (x,z)\right]},
\label{eq:Z_h_gauge_unfixed_4}
\end{equation}
where now $A_M=\{A_\mu,0\}$ indicates a 5D vector with vanishing fifth component, and $\Lambda=\tilde{\Lambda}^{-1}\circ\Sigma_1$, with $\hat{\Lambda}=\Sigma_1$ and $\bar{\Lambda}=1$. At this point, we want to make a couple of comments. First, we note that the integrand becomes completely independent of any $G_B$ element and the integral over $G_B$ (the infinite gauge redundancy) will be cancelled between the numerator and the denominator in any observable computation. Therefore, as usual, we can simply drop that factor. Second, since we have chosen the axial gauge, we do not need ghost fields to exponentiate the determinant factor. Third, the transformation parameter $\Lambda (x,z)$ can be thought of as an interpolating function between a coset element $\hat{\Lambda}$ and a trivial element $\bar{\Lambda}$. As we mentioned before, such an element exists whenever $\pi_4 (G/H) = 0$.

We finally arrive at the gauge-fixed holographic partition function for arbitrary choice of UV and IR BC:
\beq
Z^{\rm g.f.} \left[B^A\right] =  \int \mathcal{D} B^i \; \mathcal{D} \Sigma_1 (x) \mathcal{D} A_\mu (x,z) |^{\hat{A} = B^{\Sigma_1^{-1}}}_{\bar{A} :  \left[
\begin{array}{l}
\scriptscriptstyle  F_1^i = 0 \\
\scriptscriptstyle  A_1^a = 0
\end{array} \right. } 
\;  e^{i S\left[A^{\Lambda} (x,z)\right]}.  \hspace{0.5cm} \textnormal{(general UV-BC, IR-BC)}
\label{eq:Z_h_gauge_fixed_general} 
\eeq
Using this general formula, we can obtain results for special cases. Two particularly relevant ones are (i) pure Dirichlet UV-BC and pure Neumann IR-BC and (ii) pure Dirichlet UV-BC and mixed IR-BC with a $G/H_1$ symmetry breaking pattern. In the first case, we simply remove the integration over $B^i$ and $\Sigma_1$ and set $\Sigma_1 = 1$ (hence $\Lambda = 1$ as well). In the second case, while we remove the $B^i$-integral, we keep $\Sigma_1$ and $\Lambda$ as they are. We obtain\footnote{We have confirmed the following results by separate explicit computations.}
\begin{align}
&Z^{\rm g.f.} \left[B^A\right] =  \int \mathcal{D} A_\mu (x,z) |^{\hat{A} = B}_{\bar{A} : F^A=0}
\;  e^{i S\left[A (x,z)\right]},  \hspace{0.35cm} \textnormal{(D-UV-BC, N-IR-BC)}
\label{eq:Z_h_gauge_fixed_D-UV-BC_N-IR-BC}  \\
&Z^{\rm g.f.} \left[B^A\right] =  \int \mathcal{D} \Sigma_1 (x) \mathcal{D} A_\mu (x,z) |^{\hat{A} = B^{\Sigma_1^{-1}}}_{\bar{A} :  \left[
\begin{array}{l}
\scriptscriptstyle  F_1^i = 0 \\
\scriptscriptstyle  A_1^a = 0
\end{array} \right. } 
\;  e^{i S\left[A^{\Lambda} (x,z)\right]}.  \hspace{0.35cm} \textnormal{(D-UV-BC, }G/H_1\textnormal{-IR-BC)}
\label{eq:Z_h_gauge_fixed_D-UV-BC_G/H-IR-BC} 
\end{align}

\section{Unbroken symmetry and `t Hooft anomaly matching}
\label{sec:anomaly_inflow_wo_SSB}

Having developed the necessary formalism to study gauge theories in a slice of $\textnormal{AdS}_5$ holographically, we now turn to the holography of anomaly inflow. In this section, we focus on the case where the IR-BC is purely Neumann. In its 4D dual CFT, this corresponds to the global symmetry (either weakly gauged or not) unbroken by the vacuum condensate. The case of mixed IR-BC with the breaking pattern $G/H_1$ will be the subject of the next section.

\subsection{Purely global symmetry}

We first study the case with pure Dirichlet UV-BC. The dual CFT then has a purely global symmetry $G$, without any gauging. 
In particular, the relevant partition function is eq.~(\ref{eq:Z_h_gauge_fixed_D-UV-BC_N-IR-BC}) and the $B^A$ associated with all $T^A \in \mathbf{g}$ are non-dynamical background fields. 

In order to study the (in)variance of the theory, we check how the partition function transforms as we vary the source fields (see appendix~\ref{appendix:symmetry_of_Z_from_source_variation}). Considering an infinitesimal transformation $g \approx 1 - \alpha$, the partition function transforms as
\beq
\begin{aligned}
Z^{\rm g.f.} \left[ \left( B^A \right)^\alpha \right] &= \int \mathcal{D} A_\mu (x,z) |^{\hat{A} = B^\alpha}_{\bar{A} : F^A=0}
\;  e^{i S_0[A] + i S_{\rm CS} [A]} \\
&= \int \mathcal{D} A_\mu (x,z) |^{\hat{A} = B}_{\bar{A} : F^A=0}
\;  e^{i S_0[A] + i S_{\rm CS} [A^\alpha]},
\end{aligned}
\eeq
where we made a change of variable $A \to A^\alpha$ and used the $G$-invariance of the gauge action $S_0$ and of the IR-BC. 
In addition, given a UV-localized group element $g (x) \approx 1 - \alpha (x)$ acting on the source $B$, we extended it to a 5D one as $\alpha (x,z) = \alpha(x)$.\footnote{In the case of $G/H_1$ discussed in section~\ref{sec:SSB_WZW}, an extra subtlety appears regarding the extension of a 4D local gauge group element to a 5D one.}
From the second line of this equation, it is clear that the (in)variance of the partition function when the UV-BC is purely Dirichlet is fully determined by the transformation of the CS action.
Using eq.~(\ref{eq:S_CS_in_terms_of_omega_5}) and
\beq
\int_{\rm 5D} \delta_\alpha \omega_5^{(0)} (A) = \int_{\rm 5D} d \omega_4^{(1)} (\alpha, A)
\eeq
we obtain
\beq
Z^{\rm g.f.} \left[ \left( B^A \right)^\alpha \right] = e^{- ic \int_{\rm UV} \omega_4^{(1)} (\alpha, B)} \int \mathcal{D} A_\mu (x,z) |^{\hat{A} = B}_{\bar{A} : F^A=0} \; e^{i S_0 [A]} \; e^{ic \int_{\rm IR} \omega_4^{(1)} (\alpha, \bar{A})} \; e^{i S_{\rm CS} [A]}.
\label{eq:anomaly_UV-D-BC_N-IR-BC_both}
\eeq
Here, since the UV brane-localized variance term is independent of $A_\mu$, we factor it out of the path integral.  
We also observe that the partition function is not invariant under the transformation. Before we proceed any further, however, we first need to discuss one problem. That is, under the transformation, the integrand picks up an IR brane-localized variance term. Recalling that the bulk gauge symmetry $G$ is unbroken at the IR brane, such an IR brane-localized variance is not acceptable.\footnote{In fact, there is an alternative to this view point. Since the 5D gauge theory is intrinsically non-renormalizable, it is at best an effective field theory. An effective gauge theory with non-vanishing gauge anomaly can still be consistently quantized below a cut-off scale and the associated cut-off scale can be estimated by the knowledge of the anomaly~\cite{Preskill:1990fr}. A study of a 5D $U(1)$ gauge theory along this line was presented in~\cite{Gripaios:2007tk}.} In order to remedy this issue, we introduce an IR-localized effective action, $\Gamma_{\rm IR} \left[\bar{A}\right]$, which under an infinitesimal transformation $A^\alpha$ shifts by an opposite anomaly factor to cancel the bulk-generated anomaly factor:
\beq
e^{i \Gamma_{\rm IR} \left[\bar{A^\alpha}\right]}   = e^{- ic \int_{\rm IR} \omega_4^{(1)} (\alpha, \bar{A})} e^{i \Gamma_{\rm IR} \left[\bar{A}\right]}.
\eeq
Such an effective action may be obtained by first introducing an anomalous set of 4D Weyl fermions charged under $G$ and localized on the IR brane. Integrating out the fermions then generates $\Gamma_{\rm IR} \left[\bar{A}\right]$:
\beq
e^{i \Gamma_{\rm IR} \left[\bar{A}\right]} \equiv \int \mathcal{D} \psi \mathcal{D} \bar{\psi} \; e^{i S_{\rm IR} \left[\bar{A}, \psi, \bar{\psi}\right]}.
\eeq
In this case, however, the coefficient $c$ cannot be arbitrary, and in fact, must be an integral multiple of the coefficient of the chiral anomaly due to a single Weyl fermion. This leads to the \emph{quantization} condition for the CS level $c$. This issue will be discussed in section~\ref{sec:quantization}.

The UV-localized variance term, on the other hand, is perfectly fine: $G$ is completely broken on the UV brane. 
In terms of the modified partition function $\tilde{Z}^{\rm g.f.}$ with $\Gamma_{\rm IR}$ inserted, the transformation rule for the holographic partition function is therefore given by
\beq
\begin{gathered}
 \tilde{Z}^{\rm g.f.} \left[ B^A \right] = \int \mathcal{D} A_\mu (x,z) |^{\hat{A} = B}_{\bar{A} : F^A=0}
\;  e^{i S_0[A] + i S_{\rm CS} [A]} \; e^{i \Gamma_{\rm IR} \left[\bar{A}\right]},   \\
 \tilde{Z}^{\rm g.f.}  \left[ \left( B^A \right)^\alpha \right] = e^{- ic \int_{\rm UV} \omega_4^{(1)} (\alpha, B)} \; \tilde{Z}^{\rm g.f.} \left[ B^A \right].
\label{eq:anomaly_D-UV-BC_N-UV-BC}
\end{gathered}
\eeq

In order to obtain the dual 4D CFT interpretation, we now view $\tilde{Z}^{\rm g.f.} \left[B^A\right]$ as the partition function of a 4D CFT with classical source $\left(B^A\right)_\mu$ coupled to the CFT current operators $J^{\mu}$. The phase factor in eq.~(\ref{eq:anomaly_D-UV-BC_N-UV-BC}) shows that the global symmetry $G$ of the CFT is anomalous. This anomalous global symmetry $G$ is unbroken by confinement, a fact dual to the pure Neumann IR-BC. Furthermore, the appearance of both the UV- and IR-localized anomaly factors in eq.~(\ref{eq:anomaly_UV-D-BC_N-IR-BC_both}), with same magnitude but opposite signs, encodes what we might call a \emph{`t Hooft anomaly matching}. In order to see this explicitly, as is usually done, let us first weakly gauge $G$. This is done by switching the UV-BC to be purely Neumann. Under this change, what we called the source fields $B$ before now turn into dynamical fields, and the UV-localized variance needs to be cancelled just like we did for the IR-localized anomaly factor. We proceed as we did for the IR-localized anomaly term, by introducing a UV-localized effective action $\Gamma_{\rm UV} [B]$, which again may be obtained by integrating out UV-localized 4D Weyl fermions. In order to achieve gauge anomaly cancellation, we require
\beq
e^{i \Gamma_{\rm UV} [B^\alpha]} =  e^{+ ic \int_{\rm UV} \omega_4^{(1)} (\alpha, B)} \; e^{i \Gamma_{\rm UV} [B]}.
\eeq
The full partition function then becomes
\beq
\tilde{Z}^{\rm g.f.}  = \int \mathcal{D} B^A \; e^{i \Gamma_{\rm UV} [B]} \; \mathcal{D} A_\mu (x,z) |^{\hat{A} = B}_{\bar{A} : F^A=0} 
\;  e^{i S_0[A] + i S_{\rm CS} [A]} \; e^{i \Gamma_{\rm IR} \left[\bar{A}\right]}.
\eeq
As expected, the partition function is independent of any $B$ fields, and the symmetry property of the theory is tested by making a change of variable (or field redefinition) in the form of a $G$-transformation and check whether it results in an anomalous change to the original theory or not.
 
The 4D interpretation goes as follows. The 4D confining gauge theory (in fact deformed CFT) has a weakly gauged symmetry $G$. In addition to the CFT preons, there is an anomalous set of external fermions (UV-localized fermions that induce $\Gamma_{\rm UV}$) charged under $G$. While CFT and external sector are not individually $G$-anomaly free, these two contributions, nonetheless, cancel, making the gauging of $G$ legal. Importantly, these external fermions need not couple to the confining CFT gauge force, and they are spectator fermions. Along the renormalization group (RG) flow, while the chiral anomaly does not get renormalized, the CFT sector undergoes confinement. By assumption, $G$ is not broken, and the original anomaly of the CFT preons should be reproduced by a spectrum of massless composite fermions. 
These composite fermions are the IR brane-localized fermions we introduced in 5D to cancel the IR-localized variance. 
For these reasons, the choice of Neumann IR-BC and the resulting requirement of cancelling the IR-localized variance term by 4D fermions on the IR brane is the holographic realization of `t Hooft anomaly matching.

One perhaps interesting feature deduced from the above discussion is that the anomaly inflowed from the bulk CS theory must be the one that has vanishing mixed anomaly between global $G$ and confining gauge group $G_s$ of the CFT. In order to make this point clear, we may consider a $U(1)$ CS theory in the bulk. The variances induced on the boundaries are dual to a $U(1)^3$ anomaly of the global symmetry. If this global $U(1)$ current had a Adler-Bell-Jackiw (ABJ) type anomaly~\cite{Adler:1969gk, Bell:1969ts} with the confining gauge force of the CFT, then the spectators would be necessarily coupled to the strong interaction as well, and as a result, `t Hooft argument for the anomaly matching would not hold. However, we have seen that, with Neumann IR-BC, anomaly inflow by the bulk CS action always comes with an IR-localized variance term in addition to the UV variance term: `t Hooft anomaly matching is automatically at play. As we discuss in section~\ref{subsec:no_SSB_partial_gauging}, this fact holds even if $G$ is weakly gauged.

\subsection{Partially gauged symmetry}
\label{subsec:no_SSB_partial_gauging}

In this section, we consider the anomaly inflow with a mixed UV-BC in which a subgroup $H_0 \subset G$ takes a Neumann BC, while the coset $G/H_0$ has a Dirichlet BC. In order to make the discussion as concrete as possible, and also in part to demonstrate the usage of the formalism we develop in appendix~\ref{sec:Cartan_homotopy_formula}, we consider a product group $G= G_1 \times G_2$, where $G_1 = U(1)$ and $G_2$ is any compact simple Lie group. We choose UV-BC such that one factor group takes Neumann while the other takes Dirichlet BC. This choice is interesting because it admits a non-trivial mixed CS action, hence a mixed anomaly interpretation in the dual 4D picture. A lot of qualitative features we describe below apply to more general cases and an explicit analysis with arbitrary choice of $G \to H_0$ can be achieved straightforwardly.

In the 4D dual description, the CFT has a global $G= G_1 \times G_2$ symmetry and one factor group (either $G_1$ or $G_2$) is weakly gauged. We present both cases, one with gauged $G_1$ and the other with gauged $G_2$. With pure Neumann IR-BC, none of these symmetries are broken at the confinement scale. The anomaly inflow from the mixed CS action will get a 4D interpretation in terms of $G_1$-$G_2$ mixed anomaly, while inflow by the pure CS actions corresponds to pure $G_1$ and/or $G_2$ anomalies.

Let us choose $G_2 = SU(2)$. Any other choice of $G_2$ will require a very similar discussion.\footnote{One exceptional property of the group $SU(2)$ is that it has vanishing $d^{abc} \propto {\rm Tr} [ T^a \{ T^b, T^c \} ]$ and the pure $SU(2)$ CS action vanishes identically. This feature is dual to the fact that there is no perturbative $SU(2)$ anomaly in 4D.} The bulk CS action consists of two contributions: pure $U(1)$ and mixed $U(1) \text{-} SU(2)$. The \emph{form} of the mixed CS action can be obtained by first embedding $U(1)$ and $SU(2)$ into a simple compact Lie group $G_{\rm GUT}$. Once this is done, then the GUT gauge field $A$ can be written as a sum $A = V + W$ in terms of the $U(1)$ gauge field $V$ and the $SU(2)$ gauge field $W$. The mixed CS action can be read off from the CS action of $A$~\cite{Bai:2009ij}. In terms of the \emph{canonical} $\omega_5^{(0)} (A)$, after a couple of integrations by parts, we get (including the pure $U(1)$ CS action)
\beq
S_{\rm CS} [V,W] = c_1 \int_{\rm 5D} \; V dV dV + c_{12} \int_{\rm 5D} \; 3 {\rm Tr} \left[ V F_W^2 \right] +  d {\rm Tr} \left[ \left(2 V W dW + \frac{3}{2} V W^3 \right) \right],
\label{eq:U(1)-SU(2)_CS}
\eeq
where $F_W = d W + W^2$ is the $SU(2)$ field strength 2-form. Notice that the last term, obtained as a result of integration by parts, is a brane-localized term. The virtue of this form for the CS action is that the bulk CS actions are manifestly $SU(2)$-invariant and any non-trivial $SU(2)$ transformations are from the boundary terms. It may be worth mentioning that a priori the two CS levels $c_1$ and $c_{12}$ are independent and are subject to separate quantization conditions (see section~\ref{sec:quantization}). 

Under $U(1)$ and $SU(2)$ transformations, the CS action changes as
\bea
&& \delta_{\alpha_1} S_{\rm CS} = c_1 \int_{\rm IR} \; \alpha_1 d\bar{V} d \bar{V} + c_{12} \int_{\rm IR} \; \left( 3 {\rm Tr} \left[ \alpha_1 F_{\bar{W}}^2 \right] - {\rm Tr} \left[ \alpha_1 \left( 2 d \bar{W} d \bar{W} + \frac{9}{2} d \bar{W} \bar{W}^2 \right) \right] \right) \nonumber \\
&& \phantom{{}\delta_{\alpha_1} S_{\rm CS}={}} - c_1 \int_{\rm UV} \; \alpha_1 d\hat{V} d \hat{V} - c_{12} \int_{\rm UV} \; \left( 3 {\rm Tr} \left[ \alpha_1 F_{\hat{W}}^2 \right] - {\rm Tr} \left[ \alpha_1 \left( 2 d \hat{W} d \hat{W} + \frac{9}{2} d \hat{W} \hat{W}^2 \right) \right] \right), \nonumber \\
&& \label{eq:SU(2)_U(1)_U(1)_transf} \\
&& \delta_{\alpha_2} S_{\rm CS} = c_{12} \int_{\rm IR} \; {\rm Tr} \left[2 \bar{V} d \alpha_2 \left( d \bar{W} - 2 \bar{W}^2 \right) + \frac{9}{2} \bar{V} \left( d \alpha_2 \bar{W}^2 \right) \right] \nonumber \\
&& \phantom{{}\delta_{\alpha_2} S_{\rm CS}={}} - c_{12} \int_{\rm UV} \; {\rm Tr} \left[2 \hat{V} d \alpha_2 \left( d \hat{W} - 2 \hat{W}^2 \right) + \frac{9}{2} \hat{V} \left( d \alpha_2 \hat{W}^2 \right) \right].
\label{eq:SU(2)_U(1)_SU(2)_transf}
\eea
Next we discuss two cases, one with gauged $G_1$ and the other with gauged $G_2$ in turn.

\subsubsection{$G=G_1 \times G_2$ with gauged $G_1$}

We first consider the case with Neumann UV-BC for $U(1)$ and Dirichlet UV-BC for $SU(2)$. The relevant partition function is 
\beq
Z^{\rm g.f.} \left[B^A\right] = \int \mathcal{D} B^i \; \mathcal{D} A_\mu \vert^{\hat{A}=B}_{\bar{A}: F^A=0} \; e^{i S_0 + i S_{\rm CS}},
\eeq
with $S_{\rm CS}$ given in eq.~(\ref{eq:U(1)-SU(2)_CS}). In this case, $B^i = \hat{V}$ and $B^a = \hat{W}$. 
%
%
%
Since the full $G = G_1 \times G_2$ is unbroken on the IR brane, we need to cancel the anomaly factors there.
In the 4D dual description, the $c_1$-term in eq.~(\ref{eq:SU(2)_U(1)_U(1)_transf}) corresponds to the $U(1)^3$ anomaly, while the $c_{12}$-terms in eq.~(\ref{eq:SU(2)_U(1)_U(1)_transf}) and~(\ref{eq:SU(2)_U(1)_SU(2)_transf}) represent the mixed anomaly. One way to eliminate the IR-localized variance is to introduce IR-localized Weyl fermions charged under both $U(1)$ and $SU(2)$. The requirement is that this set is anomalous in such a way that their $U(1)^3$ and mixed anomalies cancel the CS-generated variance terms. 
In order to present another possibility, however, we take a slightly different path. 

As discussed in detail in appendix~\ref{sec:Cartan_homotopy_formula}, we can add a local counter terms $d B_4 (A_0, A_1)$ to the bulk CS term in such a way that the CS action becomes invariant under an $H_0 \subset G$ transformation. In particular, this works well for the product group (see appendix~\ref{subsec:shifted_anomaly_local_counter_term}). Applying this to the mixed CS action, in the current example, we take $A_0 = V$ and $A_1 = A = V + W$. The shifted mixed CS action $\omega_5^{(0)} \to \tilde{\omega}_5^{(0)} (V, A) = \omega_5^{(0)} (A) - \omega_5^{(0)} (V) + d B_4 (V, A)$ is then invariant under a $U (1)$ transformations. To be more precise, we first note that $\omega_5^{(0)} (A)$ contains the pure $U(1)$ CS action as well as the mixed CS action. Using a short notation for the mixed CS action as $\omega_5^{(0)} ({\rm mixed}) = \omega_5^{(0)} (A) - \omega_5^{(0)} (V)$, what we really do is
\beq
\begin{aligned}
S_{\rm CS} & =  c_1 \int \omega_5^{(0)} (V) + c_{12} \int \omega_5^{(0)} ({\rm mixed})  \\
 &\to  c_1 \int \omega_5^{(0)} (V) + c_{12} \int \tilde{\omega}_5^{(0)} (V,A) \\
& =  c_1 \int \omega_5^{(0)} (V) + c_{12} \int \left[ \omega_5^{(0)} ({\rm mixed}) + d B_4 (V,A) \right].
\label{eq:U(1)_SU(2)_shifted_mixed_CS}
\end{aligned}
\eeq
Notice that in $\tilde{\omega}_5^{(0)} (V,A)$ there is a cancellation between two $U(1)$ CS actions, and effectively the procedure is equivalent to adding a local counter terms $d B_4$ to the original mixed CS action. An important property we recover is that this shifted CS action is invariant under a $U(1)$ transformation. We will write the shifted CS action as $\tilde{S}_{\rm CS} = S_{\rm CS} + S_{\rm CT}$, where $S_{\rm CT}$ is the action for the counter terms. In order to show the $U(1)$-invariance more explicitly, we first note that from eq.~(\ref{eq:B_4}) the explicit form of the counter term $B_4$ in this case is given by
\beq
\begin{gathered}
S_{\rm CT} = c_\textnormal{CT} \int_{\rm 5D} \; d B_4 (V,A), \\
B_4 (V,A) = {\rm Tr} \left[ V W dW + \frac{1}{2} V W^3 \right].
\end{gathered}
\eeq
One may notice that these two terms in $B_4$ are exactly the same as the boundary terms in eq.~(\ref{eq:U(1)-SU(2)_CS}), only the relative coefficients differ. Also, eventually, $c_\textnormal{CT} = c_{12}$ as is evident from eq.~(\ref{eq:U(1)_SU(2)_shifted_mixed_CS}). Here, we use a separate notation temporarily to make one important point below. Explicit computations show that with counter term added, the shifted CS action transforms according to
\bea
&& \delta_{\alpha_1} \tilde{S}_{\rm CS} = c_1 \int_{\rm IR} \; \alpha_1 d \bar{V} d \bar{V} + c_{12} \int_{\rm IR} \; \left( 1 - \frac{c_\textnormal{CT}}{c_{12}} \right) {\rm Tr} \left[ \alpha_1 F_{\bar{W}}^2 \right] \nonumber \\
&& \phantom{{}\delta_{\alpha_1} \tilde{S}_{\rm CS}={}} - c_1 \int_{\rm UV} \; \alpha_1 d \hat{V} d \hat{V} - c_{12} \int_{\rm UV} \; \left( 1 - \frac{c_\textnormal{CT}}{c_{12}} \right) {\rm Tr} \left[ \alpha_1 F_{\hat{W}}^2 \right], \\
&& \delta_{\alpha_2} \tilde{S}_{\rm CS} = c_{12} \int_{\rm IR} \; \left( 2+\frac{c_\textnormal{CT}}{c_{12}} \right) {\rm Tr} \left[ \bar{V} d \alpha_2 d \bar{W} \right] - c_{12} \int_{\rm UV} \; \left( 2+\frac{c_\textnormal{CT}}{c_{12}} \right) {\rm Tr} \left[ \hat{V} d \alpha_2 d \hat{W} \right]. \nonumber \\
\eea
It is observed that the mixed anomaly terms are such that in units of $c_{12}$ the ``sum'' of $U(1)$ and $SU(2)$ variations is fixed, $( 1 - c_\textnormal{CT}/c_{12} ) + (2 + c_\textnormal{CT}/c_{12} ) = 3$, regardless of the size of the counter term. In contrast, the ``difference'' is not fixed, and in fact is proportional to the size of the counter term~\cite{Preskill:1990fr}. We also confirm that with $c_\textnormal{CT} = c_{12}$, the shifted \emph{mixed} CS action is indeed invariant under $U(1)$ transformations: all the mixed anomaly is attributed to the $SU(2)$ currents.

We now introduce a set of IR-localized Weyl fermions charged under both $U(1)$ and $SU(2)$ such that their chiral anomaly cancels the CS-induced variance terms. Equivalently, we add $\Gamma_{\rm IR} \left[\bar{V}, \bar{W}\right]$ which transforms as
\beq
\begin{gathered}
 e^{i \Gamma_{\rm IR} \left[\bar{V}^{\alpha_1}, \bar{W}\right]} = e^{- i c_1 \int_{\rm IR} \; \alpha_1 F_{\bar{V}}^2} \; e^{i \Gamma_{\rm IR} \left[\bar{V}, \bar{W}\right]}, \\
 e^{i \Gamma_{\rm IR} \left[\bar{V}, \bar{W}^{\alpha_2}\right ]} = e^{- i c_{12} \int_{\rm IR} \; 3 {\rm Tr} \left[ \bar{V} d \alpha_2 d \bar{W} \right]}  \; e^{i \Gamma_{\rm IR} \left[\bar{V}, \bar{W}\right]}.
\end{gathered}
\eeq
Considering the UV-localized variance term, since only the $U(1)$ factor takes Neumann BC, we only need to cancel the pure $U(1)^3$ variance term. Hence, on the UV brane, we add a set of Weyl fermions charged under $U(1)$ only. The UV brane-localized effective action, upon integrating out these fermions, is then required to transform as
\beq
e^{i \Gamma_{\rm UV} \left[\hat{V}^{\alpha_1}\right]} = e^{+ i c_1 \int_{\rm UV} \; \alpha_1 F_{\hat{V}}^2} \; e^{i \Gamma_{\rm UV} \left[\hat{V}\right]}. 
\eeq

Moving on to the 4D dual interpretation, at the UV scale, the theory consists of a CFT sector and a set of external fermions. The CFT has a global symmetry group $G = SU(2) \times U(1)$, of which the $U(1)$ factor is weakly gauged. The external fermions are charged under the $U(1)$ gauge force. The $U(1)^3$ gauge anomaly is cancelled between the two contributions from the CFT and the external sector. There is a non-vanishing $U(1) \text{-} SU(2)$ mixed anomaly of ABJ type, which comes only from the CFT sector. Naively, depending on the UV regulator, the mixed anomaly can be shared among gauge and global currents. In particular, if the gauge current is anomalous, we have an issue with gauging the $U(1)$ factor. However, we added local counter terms proportional to $B_4 (V, A)$, so that we moved all the mixed anomaly to the global $SU(2)$ currents. In this way, the gauged $U(1)$ symmetry is free of any mixed anomalies. Thanks to this feature, the external fermions need not carry the global $SU(2)$ quantum numbers. This is an analog of what occurs in QCD: there the anomaly computed from the Feynman diagrams (\emph{consistent anomaly}) results in the non-conservation of both vector and axial-vector currents. However, by adding an appropriate counter term (\emph{Bardeen's counter term}), the vector current becomes conserved and all the mixed anomaly is moved to the axial-vector current (\emph{covariant anomaly})~\cite{Bardeen:1969md}. 

As the theory RG runs to the IR scale, the CFT sector confines and the anomalies are matched by massless composite fermions. Notice that in the current example, the $U(1)$ factor is physically gauged. Nevertheless, anomaly matching arguments go through since $U(1)$ is not a confining force. As for the $SU(2)$ part, as usual, we formally weakly gauge it, and introduce extra external spectator fermions to cancel the mixed anomaly. This mixed anomaly is also reliably reproduced by the massless composite fermions in the IR. Our holographic study of anomaly inflow, therefore, shows that in a confining gauge theory, the anomaly associated with a weakly gauged symmetry in the UV (i.e.~$U(1)$ gauge anomaly carried by the composite sector) as well as the ABJ anomaly (i.e.~mixed $U(1) \text{-} SU(2)$) are matched by the composite spectrum in the IR.

\subsubsection{$G=G_1 \times G_2$ with gauged $G_2$}

In this case, we take $A_0 = W$ and $A_1 = A$. The counter term $B_4 (W, A)$ is similarly given by
\beq
B_4 (W, A) = - {\rm Tr} \left[ 2 V W d W + \frac{3}{2} V W^3 \right].
\eeq
This exactly cancels the boundary terms in eq.~(\ref{eq:U(1)-SU(2)_CS}) and the shifted CS action becomes manifestly $SU(2)$ invariant. Under a $U(1)$ transformation, we get
\beq
\delta_{\alpha_1}\tilde{S}_{\rm CS} = c_1 \int_{\rm IR} \; \alpha_1 F_{\bar{V}}^2 + c_{12} \int_{\rm IR} \; 3 {\rm Tr} \left[ \alpha_1 F_{\bar{W}}^2 \right] - ({\rm UV}),
\eeq
where the UV-localized variance terms are obtained from the IR-localized terms with the replacement $\bar{V}, \bar{W} \to \hat{V}, \hat{W}$. In this case, there is no gauge anomaly. All the anomalies are on the global current, either pure global $U(1)^3$ or ABJ-type mixed anomalies. On the IR, since $G$ is preserved, we again need to include brane-localized fermions. On the UV, on the other hand, thanks to the counter term $B_4$, no $SU(2)$ variance term shows up and there's no need to add anything. 

The 4D interpretation is straightforward. The CFT has a global symmetry group $G = SU(2) \times U(1)$ and $SU(2)$ is weakly gauged. While there are global $U(1)^3$ and mixed anomalies, a proper counter term is added in such a way that the gauged $SU(2)$ is free of any mixed anomaly. We can again formally weakly gauge the $U(1)$ part and add spectator fermions. In the IR, when the CFT confines, composite fermions achieve `t Hooft anomaly matching, the 4D dual of IR-localized fermions. Once again, our holographic study of anomaly inflow indicates that an ABJ anomaly in the UV, when the gauged external legs are weakly interacting, is matched by the composite spectrum in the IR. While in the previous section it was $U(1)$ that was weakly gauged, here it is a non-Abelian group, namely $SU(2)$, and the same analysis can be performed with any non-Abelian group.

\section{Spontaneously broken symmetry and Wess-Zumino-Witten action}
\label{sec:SSB_WZW}

In this section, we consider the possibility that the IR-BC breaks $G$ down to $H_1 \subset G$. In its holographic 4D CFT dual, this corresponds to the spontaneous breaking of the symmetry group $G$ down to $H_1$ by confinement. If the UV-BC is Dirichlet for all $G$, the bulk gauge group is dual to a \emph{global} symmetry of the 4D CFT. On the other hand, choosing Neumann UV-BC for all $G$ corresponds to a \emph{weakly gauged} symmetry. A slightly less trivial case can be analyzed by choosing Dirichlet UV-BC for some of the generators, and Neumann UV-BC for the rest. For instance, if we choose Neumann UV-BC for a subgroup $H_0 \subset H_1$, and Dirichlet for the rest, the dual picture is that of a CFT with global symmetry $G$ spontaneously broken to $H_1$ by vacuum condensate, and a subgroup $H_0 \subset H_1$ is weakly gauged. This is very much like what happens in QCD. There, $G$ is the chiral symmetry $G = SU(3)_L \times SU(3)_R$, which is broken to $H_1 = SU(3)_V$ by the quark condensate. Moreover, $U(1)_{\rm EM} \subset H_1$ is weakly gauged.

Using the result of section~\ref{sec:holo_Z}, we start with eq.~(\ref{eq:Z_h_gauge_fixed_general}), which we report below again for convenience:
\beq
Z^{\rm g.f.} \left[B^A\right] = \int \mathcal{D} B^i \; \mathcal{Z} \left[B^A\right] = \int \mathcal{D} B^i \; \mathcal{D} \Sigma_1 (x) \mathcal{D} A_\mu (x,z) |^{\hat{A} = B^{\Sigma_1^{-1}}}_{\bar{A} :  \left[
\begin{array}{l}
\scriptscriptstyle  F_1^i = 0 \\
\scriptscriptstyle  A_1^a = 0
\end{array} \right. } 
\;  e^{i S\left[A^{\Lambda} (x,z)\right]}, 
\label{eq:SSB_Z_gf_general} \\
\eeq
where $\hat{\Lambda} =  \Sigma_1$ and $\bar{\Lambda} = 1$. Before we delve into a detailed discussion of the two separate cases (Dirichlet UV-BC vs Neumann UV-BC) let us study the gauge transformation properties of $\mathcal{Z} \left[B^A\right]$. Once this is understood, it's easier to focus on a specific case.

The action consists of the gauge kinetic term, $S_0$, and of the CS action, which for now we set to its \emph{canonical} version $S_{\rm CS} = c \int \omega_5^{(0)} (A)$. Also, for the sake of simplicity, we will just write $\bar{A}$ for the IR-BC. For example, IR-BC after a gauge transformation by $h \in H$ will be denoted as $\bar{A}^h$, and this means $(F_1^h)^i = 0$ for the unbroken generators and $(A_1^h)^a = 0$ for the broken ones. Under $\hat{g} \in G$ on the UV brane, $\mathcal{Z} \left[B^A\right]$ transforms as
\beq
\begin{aligned}
\mathcal{Z} \left[ \left( B^A \right)^{\hat{g}} \right]  &= \int \mathcal{D} \Sigma_1 \int \mathcal{D} A_\mu |^{\hat{A} = B^{\Sigma_1^{-1} \circ \hat{g}}}_{\bar{A}}
\; e^{i S_0\left[A^\Lambda\right] + i S_{\rm CS}\left[A^\Lambda\right]} \\
&= \int \mathcal{D} \Sigma_1^{\hat{g}} \int \mathcal{D} A_\mu |^{\hat{A} = B^{(\Sigma^{\hat{g}})^{-1} \circ \hat{g}}}_{\bar{A}} \; e^{i S_0 \left[A^{\Lambda^g}\right] + i S_{\rm CS} \left[A^{\Lambda^g}\right]} \\
&= \int \mathcal{D} \Sigma_1 \int \mathcal{D} A_\mu |^{\hat{A} = B^{h(\hat{g}, \Sigma) \circ \Sigma^{-1}}}_{\bar{A}} \; e^{i S_0 \left[A^{\Lambda^g}\right] + i S_{\rm CS} \left[A^{\Lambda^g}\right]}.
\end{aligned}
\eeq
In the second line, we made a change of integration variable $\Sigma_1 \to \Sigma_1^{\hat{g}}$, and used the fact that the integration measure on $G/H_1$ is the left invariant Haar measure to get the third line. 

We mention that we extend a given $\hat{g} (x) \approx 1 - \hat{\alpha} (x)$ on the UV brane to 5D such that $g(x, z_0) = \hat{g}$ and $g (x, z_1) = \bar{g} \in H_1$~\cite{Panico:2007qd}. The reason for the latter condition is simply that on the IR brane $H_1$ is the only unbroken gauge group. 
Such an extension of gauge element exists provided $\pi_4 (G/H)$ is trivial, which we assume in the paper. This can be understood by first decomposing $g = \Sigma_1 h, \; \Sigma_1 \in G/H_1, \; h \in H_1$, and realizing that the desired extension is equivalent to the deformation that takes a coset element on the UV brane into a trivial element, i.e.~$H_1$-element, on the IR brane. 

In order to make the overall transformation more manifest, next we make a change of variable: $A \to A^{h (g, \Sigma)}$. The partition function becomes
\beq
\begin{aligned}
\mathcal{Z} \left[ \left( B^A \right)^{\hat{g}} \right] &= \int \mathcal{D} \Sigma_1 \int \mathcal{D} A_\mu |^{\hat{A} = B^{\Sigma_1^{-1}}}_{\bar{A}} \; e^{i S_0 [A] + i S_{\rm CS} \left[A^{\Lambda^g \circ h}\right]}  \\
&= \int \mathcal{D} \Sigma_1 \int \mathcal{D} A_\mu |^{\hat{A} = B^{\Sigma_1^{-1}}}_{\bar{A}} \; e^{i S_0 [A] + i S_{\rm CS} \left[A^{g \circ \Lambda}\right]} ,
\end{aligned}
\eeq
where we used the $H_1$-invariance of the IR-BC and the $G$-invariance of $S_0$. We also used $g \Lambda = \Lambda^g h (g, \Lambda)$ to rewrite the argument of the CS action. Hence, we see that the theory is not invariant under a given $g \in G$ transformation, and in particular, its non-invariance comes from the bulk CS action. In order to understand the form of the transformation, it is sufficient to study the infinitesimal version. Under $\hat{g} \approx 1 - \hat{\alpha}$, we get
\beq
\begin{aligned}
\mathcal{Z} \left[ \left( B^A \right)^{\hat{g}} \right] &= \int \mathcal{D} \Sigma_1 \int \mathcal{D} A_\mu |^{\hat{A} = B^{\Sigma_1^{-1}}}_{\bar{A}} \; e^{i S_0 [A] + i S_{\rm CS} \left[A^{ \Lambda}\right] + i \delta_\alpha S_{\rm CS} \left[A^\Lambda\right]}  \\
&= e^{- i c \int_{\rm UV} \omega_4^{(1)} (\hat{\alpha}, B)} \int \mathcal{D} \Sigma_1 \int \mathcal{D} A_\mu |^{\hat{A} = B^{\Sigma_1^{-1}}}_{\bar{A}} \; e^{i S_0 [A] + i S_{\rm CS} \left[A^{ \Lambda}\right]  } \; e^{+ i c  \int_{\rm IR} \omega_4^{(1)} (\bar{\alpha}, A_h)} . 
\label{eq:G/H_gauge_transf_pure_D-UV-BC_general}
\end{aligned}
\eeq
We used that $A_\mu (x, z_0) = \hat{A}_\mu = B_\mu^{\Sigma_1^{-1}}$ and $A_\mu (x, z_1) = \bar{A}_\mu = A_h$, where $A_h$ is the restriction of the gauge field to $\mathbf{h}_1$. It may be worth mentioning that with $(A)^{g \circ \Lambda} = (A^\Lambda)^g$ we denote the whole gauge transformation of $A^\Lambda$ as a single gauge connection, i.e.~$\left(A^\Lambda\right)^g = g \left( d + A^\Lambda \right) g^{-1}$.

In order to restore 5D consistency, we need to modify the theory to remove the IR-localized variance term. This may be attained by adding an appropriately anomalous set of localized 4D fermions. 
Alternatively, the problem is solved if the subgroup $H_1$ is an anomaly-free embedding (AFE). Anomaly-free embedding means $\omega_5^{(0)} (A_h) = 0$, i.e.~the CS action (hence associated anomaly $\omega_4^{(1)}$) vanishes when the gauge field $A$ is restricted to its $H_1$ part. In what follows, we take the second path.\footnote{In fact, in the derivation of eq.~(\ref{eq:SSB_Z_gf_general}) it was already assumed that $H_1$ is AFE.}
In addition, we also promote $\omega_5^{(0)} (A) \to \tilde{\omega}_5^{(0)} (A_h, A)$ so that the bulk CS action is invariant under $H_1$ transformations.
When it comes to the UV-localized variance term, the required amendment and associated dual interpretation depends on the UV-BC.

\subsection{Purely global symmetry}
\label{subsec:SSB_purely_global}

When all of $G$ satisfies Dirichlet BC on the UV brane, from a 5D perspective, there is no induced UV surface terms as a result of a gauge transformation. Hence, once we cure the IR brane-localized non-invariance, the 5D theory is consistent. Denoting the shifted CS action as $\tilde{S}_{\rm CS} = c \int \tilde{\omega}_5^{(0)} (A_h, A)$, eq.~(\ref{eq:G/H_gauge_transf_pure_D-UV-BC_general}) becomes
\beq
\begin{aligned}
&Z^{\rm g.f.} \left[B^A\right] = \int \mathcal{D} \Sigma_1 \int \mathcal{D} A_\mu |^{\hat{A} = B^{\Sigma_1^{-1} }}_{\bar{A}}
\; e^{i S_0 [A] + i \tilde{S}_{\rm CS}\left[A^\Lambda\right]} \\
\to \;&Z^{\rm g.f.} \left[ \left( B^A \right)^{\hat{\alpha}} \right] =  e^{- i c \int_{\rm UV} \tilde{\omega}_4^{(1)} (\hat{\alpha}, B) } \; Z^{\rm g.f.} \left[B^A\right].
\label{eq:SSB_purely_global_transf}
\end{aligned}
\eeq
%
%
%
We remark that while the theory is not invariant under $\hat{\alpha} \in \mathbf{k}_1$ transformations, the anomalous phase $\tilde{\omega}_4^{(1)} (\hat{\alpha},B)$ vanishes for $\hat{\alpha} \in \mathbf{h}$ transformations.

In the dual 4D CFT, we interpret this as an anomalous global symmetry $G$ of the CFT. More precisely, the global symmetry $G$ is spontaneously broken at the confinement scale by the vacuum condensate, and while the unbroken group $H_1$ is free of anomalies (thanks to the counter term we added), the anomaly associated with $G/H_1$ is captured by the above phase factor. Since $H_1 \subset G$ is already anomaly-free, it can be gauged if wanted. In 5D, this is equivalent to the statement that we can freely switch the UV-BC to Neumann without needing extra modifications.

We have seen that if $H_1 \subset G$ is an AFE, no IR brane-localized state is needed. Given that the UV phase has a non-trivial anomaly, this raises a question about anomaly matching. 
In QCD, the chiral anomaly in the UV quark phase is matched in the IR hadronic phase by the gauged WZW term of Nambu-Goldstone bosons (NGBs)~\cite{Witten:1983tw}. 
It is then natural to ask if this feature can be seen in the current framework. After all, since we take Dirichlet UV-BC for all $G$, all of $G/H_1$ describes physical NGBs in the dual 4D CFT. In light of the discussion given in appendix~\ref{sec:WZW_review}, it can be shown that the answer to this question is yes, but with some subtleties. To see this, we first rewrite the (shifted) CS action in the partition function as
\beq
\tilde{\omega}_5^{(0)} \left(\left( A^\Lambda \right)_h, A^\Lambda \right) \to \tilde{\omega}_5^{(0)} (A_h, A) + \left( \tilde{\omega}_5^{(0)}  \left(\left( A^\Lambda \right)_h, A^\Lambda \right) - \tilde{\omega}_5^{(0)} (A_h, A) \right) = \tilde{\omega}_5^{(0)} (A_h, A) + \frac{\mathcal{L}_{\rm WZW}}{c}. 
\label{eq:G/H_UV-D-BC_WZW_term}
\eeq
First of all, the new version is trivially equal to the original CS action. In order to see how the expression in parenthesis is indeed the wanted WZW action, we note that it vanishes trivially as $\Sigma_1 \to 1$, and it depends only on the (UV) boundary value.\footnote{$d \left( \tilde{\omega}_5^{(0)}  \left(\left( A^\Lambda \right)_h, A^\Lambda \right) - \tilde{\omega}_5^{(0)} (A_h, A) \right) = \Omega_6 (A^\Lambda) - \Omega_6 (A) =0$.} Hence, it satisfies two of the three conditions for it to qualify as the WZW action. For the last requirement, i.e.~solving the anomalous Ward identity, it is sufficient to show that $\tilde{\omega}_5^{(0)} (A_h, A)$, as part of the partition function, is invariant under $G$-transformations. 
This is achieved rather easily. One just recalls that under an arbitrary $g \in G$ transformation, the change of the partition function is captured by $A \to A^{h (g, \Sigma_1)}$ in the CS terms. For $\tilde{\omega}_5^{(0)} (A_h, A)$ this corresponds to just $\tilde{\omega}_5^{(0)} ((A_h)^{h (g, \Sigma_1)}, A^{h (g, \Sigma_1)})$, which is invariant according to appendix~\ref{subsec:shifted_anomaly_local_counter_term}. Therefore, the expression in parenthesis is indeed the WZW term. 
For the same reason, in the splitting $\tilde{\omega}_5^{(0)} (A_h, A) + \mathcal{L}_{\rm WZW} / c$, the first term is invariant under any $g \in G$ transformation, and the non-invariance is completely encoded in the WZW term.

Crucially, in the partition function, e.g.~eq.~(\ref{eq:SSB_purely_global_transf}), the path integral variable $\Sigma_1 (x)$ depends only on the 4D spacetime coordinates, and upon integrating over the bulk, we get the holographic effective action which depends on $\Sigma_1 (x)$ as well as on the boundary value $B$. In particular, as we just showed, the WZW action only depends on the boundary value, and it can be taken out of the integral over $A_\mu (x,z)$:
\beq
\begin{aligned}
Z^{\rm g.f.} \left[B^A\right] &= \int \mathcal{D} \Sigma_1 (x) \; e^{i S_{\rm WZW} \left[B^A, \Sigma_1\right]}  \int \mathcal{D} A_\mu |^{\hat{A} = B^{\Sigma_1^{-1} }}_{\bar{A}}  \; e^{i S_0 [A] + i \tilde{S}_{\rm CS} [A]}   \\
&= \int \mathcal{D} \Sigma_1 (x) \; e^{i S_{\rm WZW} \left[B^A, \Sigma_1\right]} \; e^{i S_{\rm h,0} \left[B^A, \Sigma_1\right]}.
\label{eq:SSB_purely_global_S_WZW_S_ho}
\end{aligned}
\eeq
In the above, $S_{\rm h,0}$ is the holographic effective action obtained by integrating out the bulk degrees of freedom. With pure Dirichlet UV-BC, this action describes the chiral perturbation theory (see for example~\cite{Hirn:2005nr}) of massless NGBs. The physical NGB fields $\Sigma_1 (x)$ correspond to the Wilson line of the zero mode $A_5$, and hence constitute the low energy degrees of freedom. From this discussion, therefore, it is clear that, in the deep IR after integrating out all massive hadronic states (i.e.~integrating out the bulk in 5D), the anomaly of the global symmetry is maintained by the WZW term of NGBs.

We wish to emphasize the differences of the WZW action we obtained in the holographic context compared to the discussion given in appendix~\ref{sec:WZW_review} in the context of $2n$-dimensional spacetime. There, the WZW term is given as $\tilde{\omega}_{2n+1}^{(0)} (A_h, A) - \tilde{\omega}_{2n+1}^{(0)} \left((A^{\Lambda^{-1}})_h, A^{\Lambda^{-1}}\right)$, and the non-invariance comes from the first term, while the second term is invariant. Here, on the other hand, the WZW term is given by eq.~(\ref{eq:G/H_UV-D-BC_WZW_term}), and the non-invariance is from $\tilde{\omega}_5^{(0)} (A_h^\Lambda, A^\Lambda)$, the naive shifted CS term being invariant. Of course, this is in part because while in appendix~\ref{subsec:shifted_anomaly_local_counter_term} the $g \in G$ transformation is incorporated directly in terms of $g$, 
here it is effectively in terms of $h (g, \Sigma_1)$.


A slightly more explicit expression for the WZW action 
is obtained as follows. Since we assume that $H_1$ is an anomaly-free embedding (AFE) in $G$, the $d$-symbol, $d^{ijk} \propto {\rm Tr} [ \{ T^i, T^j \} T^k]$, vanishes for $H_1$. Using the properties listed in appendix~\ref{subsec:shifted_anomaly_local_counter_term}, we can write
\beq
\begin{aligned}
\omega_5^{(0)} (A^\Lambda) &= \tilde{\omega}_5^{(0)} (0, A^\Lambda) = \tilde{\omega}_5^{(0)} (\Lambda^{-1} d \Lambda, A) \\
&= \omega_5^{(0)} (A) - \omega_5^{(0)} (\Lambda^{-1} d \Lambda) + d B_4 (\Lambda^{-1} d \Lambda, A).
\end{aligned}
\eeq
To get the second equality, we used $\tilde{\omega}_{2n+1}^{(0)} ((A_h)^g, A^g) = \tilde{\omega}_{2n+1}^{(0)} (A_h, A)$ with $g = \Lambda^{-1}$.
The WZW action then is obtained to be
\beq
S_{\rm WZW} [\Sigma_1, B] / c= - \int_{5D} \omega_5^{(0)} (\Lambda^{-1} d \Lambda) + \int_{\partial (5D)} B_4 \left(\Lambda^{-1} d \Lambda, A\right) + B_4 \left((A^\Lambda)_h, A^\Lambda\right) - B_4 (A_h, A).
\eeq
We can use this to extract the pure NGB WZW terms. For this, we just set $A=0$. Since $B_{2n}$ vanishes if one or both of the arguments are set to zero (see appendix~\ref{subsec:shifted_anomaly_local_counter_term}), we get
\beq
S_{\rm WZW} [\Sigma_1, B=0] / c = - \int_{5D} \omega_5^{(0)} \left(\Lambda^{-1} d \Lambda\right) +  \int_{\partial (5D)} B_4 ((\Lambda d \Lambda^{-1})_h, \Lambda d \Lambda^{-1}).
\eeq 
For small $\xi (x)^a$, we expand $\Sigma_1 (x) = e^{- \xi (x)} \approx 1 - \xi^a (x) X^a$, and in particular, $\Sigma_1 d \Sigma_1^{-1} = d \xi + \mathcal{O} (\xi^2)$. The point is that infinitesimally, $\Sigma_1 d \Sigma_1^{-1} = d \xi \in \mathbf{k}$ (i.e.~$(\Lambda d \Lambda^{-1})_h=0$), and the $B_4$ term does not contribute. Using $U = \Lambda^{-1} d \Lambda \approx - d \xi$ and eq.~(\ref{eq:canonical_omega_2n+1}), straightforward steps lead to
\beq
\begin{aligned}
S_{\rm WZW} [\xi, B=0] / c &= - (-1)^5 \frac{1}{10} \int_{5D} \; {\rm Tr} \left[ (d \xi)^5 \right] + \mathcal{O} (\xi^6)  \\
&= - \frac{1}{10} \int_{\rm UV} {\rm Tr} \left[ \xi (x) (d \xi (x))^4 \right]  + \mathcal{O} (\xi^6).
\end{aligned}
\eeq
A direct comparison with the existing literature such as~\cite{Chu:1996fr} can be made by noting that for chiral symmetry, the parametrization $\Sigma = e^{- 2 i \xi}$ is used (note the factor of 2). Also, as we show in section~\ref{sec:quantization}, the overall coefficient consistent with the quantization condition is $c = \frac{\kappa}{24 \pi^2}$ with $\kappa \in \mathbb{Z}$ (see also~\cite{Hill:2006ei,Bai:2009ij}). 
Taking into account the factor $2^5$ from the difference in GB parametrization, we get
\beq
S_{\rm WZW} [\xi, B=0]  = - \frac{2}{15 \pi^2} \kappa \int_{\rm UV} {\rm Tr} \left[ \xi (x) (d \xi (x))^4 \right]  + \mathcal{O} (\xi^6), \qquad \xi = \frac{\Pi (x)}{F_\pi}.
\eeq
This agrees with~\cite{Chu:1996fr}. 

\subsection{Gauged symmetry}
\label{subsec:SSB_gauged}

In this section, we gauge some part of the group $G$, and analyze the resulting anomaly inflow. 
This is done by taking Neumann UV-BC for some of the generators of $\mathbf{g}$.
We will first consider the case in which we only gauge a subgroup  $H_0 \subset H_1$.
This is an analog of QCD, and we will obtain a \emph{gauged} version of the WZW action as the low energy effective action, as we expect. The gauged WZW action includes terms that match ABJ and `t Hooft anomalies~\cite{Witten:1983tw}. 
We then move on to analyze the most general possibility. We consider a situation in which a subgroup $H_0 \subset G$ which is not a proper subset of $H_1$ is gauged. This is a prototypical situation for dynamical symmetry breaking of the electroweak group of the SM~\cite{Weinberg:1975gm, Weinberg:1996kr}. Here, we focus on the topological terms in such a theory, and show, among other things, that the (would-be) NGBs associated with the gauged generators can be removed by means of field redefinitions.

\subsubsection{Gauging a subgroup $H_0 \subset H_1$: analog of QCD}
\label{subsubsec:SSB_gauging_subset_H_0}

Let us first consider the case in which the UV-BC for the subgroup $H_0 \subset H_1 \subset G$ is taken to be Neumann, while the rest $G/H_0$ is set to be Dirichlet.
The relevant partition function is eq.~(\ref{eq:SSB_Z_gf_general}) with the shifted CS action and $B^i_\mu \in \mathbf{h}_0$. 
As we discussed in section~\ref{subsec:SSB_purely_global}, the shifted CS action is invariant under any $H_1$ transformations.
This in turn implies that under any gauged $H_0 \subset H_1$ transformations, the shifted action is automatically invariant as well.
Therefore, we do not need to add any UV brane-localized effective action: with $\omega_5^{(0)} (A) \to \tilde{\omega}_5^{(0)} (A_h, A)$ the 5D theory is consistent.
Using eq.~(\ref{eq:G/H_gauge_transf_pure_D-UV-BC_general}), the change in the partition function under an infinitesimal $G/H_1$ transformation $\hat{\gamma} \in \mathbf{k}_1$ is found to be
\beq
\begin{aligned}
Z^{\rm g.f.} \left[ \left( B^a \right)^{\hat{\gamma}} \right] &= \int \mathcal{D} \left( B^i \right)^{\hat{\gamma}} \int \mathcal{D} \left( \Sigma_1 \right)^{\hat{\gamma}} \mathcal{D} A_\mu |^{\hat{A} = B^{\left( {\Sigma_1}^{\hat{\gamma}} \right)^{-1} \circ \hat{\gamma}}}_{\bar{A} } 
\;  e^{i S_0 [A] +  \tilde{S}_{\rm CS} \left[A^{\Lambda} (x,z)\right]} \\
&= \int \mathcal{D}  B^i \int \mathcal{D}  \Sigma_1  \mathcal{D} A_\mu |^{\hat{A} = B^{h (\hat{\gamma}, \Sigma_1) \circ \Sigma_1^{-1}}}_{\bar{A} } 
\;  e^{i S_0 [A] +  \tilde{S}_{\rm CS} \left[A^{\Lambda} (x,z)\right]}.
\end{aligned}
\eeq
Here, we changed the integration variables $B^i \to \left( B^i \right)^{\hat{\gamma}}$ and $\Sigma_1 \to \Sigma_1^{\hat{\gamma}}$ and used the invariance of the left-invariant Haar measure. In the second line, we used $g \Sigma_1 = \Sigma_1^g \; h (g, \Sigma_1)$. To extract the anomaly phase, we further redefine $A_\mu \to A_\mu^{h (\hat{\gamma}, \Sigma_1)}$ and get 
\beq
\begin{aligned}
Z^{\rm g.f.} \left[ \left( B^a \right)^{\hat{\gamma}} \right] &= \int \mathcal{D} B^i \; e^{- i c \int_{\rm UV} \tilde{\omega}_4^{(1)} \left( \hat{\gamma}, B^i, B^a \right)} \; \int \mathcal{D} \Sigma_1 \mathcal{D} A_\mu |^{\hat{A} = B^{\Sigma_1^{-1}}}_{\bar{A} } 
\;  e^{i S_0 [A] +  \tilde{S}_{\rm CS} \left[A^{\Lambda} (x,z)\right]}  \\
&= \int \mathcal{D} B^i \; e^{- i c \int_{\rm UV} \tilde{\omega}_4^{(1)} \left( \hat{\gamma}, B^i, B^a \right)}  \; \mathcal{Z} \left[B^A\right].
\label{eq:SSB_gauging_subgroup_H_0_transf_1}
\end{aligned}
\eeq

Alternatively, we can write this directly in terms of the WZW action, where the partition function is expressed as 
\beq
Z^{\rm g.f.} \left[B^A\right] = \int \mathcal{D} B^i \int  \mathcal{D} \Sigma_1 \; e^{i S_{\rm WZW} \left[B^A, \Sigma_1\right]} \; e^{i S_{\rm h,0} \left[B^A, \Sigma_1\right]}.
\eeq
Recall that $S_{\rm h,0}$ is the holographic effective action and is invariant under any $G$ transformation (see eq.~(\ref{eq:SSB_purely_global_S_WZW_S_ho})): any variance comes from the WZW action. Eq.~(\ref{eq:SSB_gauging_subgroup_H_0_transf_1}) turns into
\beq
Z^{\rm g.f.} \left[ \left( B^a \right)^{\hat{\gamma}} \right] = \int \mathcal{D} B^i \; e^{- i c \int_{\rm UV} \tilde{\omega}_4^{(1)} \left( \hat{\gamma}, B^i, B^a \right)} \int \mathcal{D} \Sigma_1  \; e^{i S_{\rm WZW} \left[B^A, \Sigma_1\right]} \; e^{i S_{\rm h,0} \left[B^A, \Sigma_1\right]}.
\label{eq:SSB_gauging_subgroup_H_0_transf_2}
\eeq
This form of transformation makes the following 4D dual interpretation very transparent.

In the 4D dual CFT, we have a global symmetry $G$ which is broken down to $H_1 \subset G$ by confinement. In addition, $H_0 \subset H_1$ is weakly gauged. In the strongly interacting CFT, we regulate the UV-divergences 
such that $H_1$ is anomaly-free. The advantage of such a choice is that the gauged symmetry $H_0$ is automatically anomaly-free. However, there can still be anomalies in global currents. In particular, we can have mixed anomalies among global and gauged currents as well as pure global anomalies. Thanks to the proper local counter terms added (i.e.~choice of UV-regulator), the mixed anomalies come entirely from the global currents. Under a global transformation $\hat{\gamma} \in \mathbf{k}_1$, the CFT is anomalous and the anomaly is captured by the phase factor in eq.~(\ref{eq:SSB_gauging_subgroup_H_0_transf_2}). We emphasize that the anomaly factor depends on both the classical source $B^a$ and the gauged dynamical field $B^i$. In detail, this single anomaly phase contains both pure global anomalies as well as mixed global-gauge anomalies. This of course is equivalent to the statement that the WZW action contains local operators that match chiral anomalies of the UV phase~\cite{Witten:1983tw}. For example, in QCD, the gauged WZW Lagrangian contains $\mathcal{L}_{\rm WZW} \supset n \frac{e^2}{24 \pi^2 F_{\pi}} \pi^0 F \tilde{F}$, which on one hand explains the $\pi^0 \to \gamma \gamma$ decay, and on the other hand, reproduces the ABJ anomaly when the quantization condition $n = N_c = 3$ is chosen. The QCD WZW Lagrangian also contains $\mathcal{L}_{\rm WZW} \supset - \frac{2}{3} i e \frac{n}{\pi^2 F_{\pi}^3} \epsilon^{\mu\nu\rho\sigma} A_\mu \partial_\nu \pi^+ \partial_\rho \pi^- \partial_\sigma \pi^0$, which in turn reproduces the QCD VAAA anomaly for $n=N_c =3$. Once we specify $G$, $H_1$, and $H_0$, our formalism allows explicit computations of all of these using the results contained in appendix~\ref{sec:Cartan_homotopy_formula} and {\ref{sec:WZW_review}.

\subsubsection{General analysis}

In this section, we consider the most general possibility in which $H_0 \subset G$ is gauged and $G$ is spontaneously broken to $H_1$.
The transformation of the partition function is obtained through a series of similar steps as in section~{\ref{subsubsec:SSB_gauging_subset_H_0}. The final result has the same form as eq.~(\ref{eq:SSB_gauging_subgroup_H_0_transf_1}) but the transformation parameter $\hat{\gamma}$ is not constrained to be just $\hat{\gamma} \in \mathbf{k}_1$ and instead takes any value $\hat{\alpha} \in \mathbf{g}$. There are three different kinds of transformations we need to consider separately. First, under $\hat{\beta} \in \mathbf{h}_1$, the partition function is invariant. As a result, there is no UV-brane surface terms associated with these transformations, hence no ``remedy'' is needed. Second, we can perform a pure global transformation $\hat{\gamma}$ which is not part of $\mathbf{h}_1$ ($G/(H_0 \cup H_1)$ transformations). For these transformations, the partition function changes in exactly the same manner as in section~\ref{subsubsec:SSB_gauging_subset_H_0} and in the 4D dual description we get anomalies in the global currents. Finally, we can consider $\hat{\delta} \in \mathbf{h}_0$ which is not part of $\mathbf{h}_1$ ($H_0 \setminus H_0 \cap H_1$ transformations). 
For these transformations, we get UV-localized anomaly factors, and given that these generators are not broken on the UV brane, we need to cure this problem. This is done by inserting 
\beq
e^{i \Gamma_{\rm UV} \left[B^i\right]} \equiv \int \mathcal{D} \psi \mathcal{D} \bar{\psi} \; e^{i S_{\rm UV} \left[B^i, \psi, \bar{\psi}\right]}
\eeq
into the partition function. As usual, we require it to transform so as to cancel the variance of the third kind (i.e.~$\hat{\delta}$ transformations). Specifically, $\Gamma_{\rm UV} \left[B^i\right]$ is invariant under both $H_1$ and $G / (H_0 \cup H_1)$ transformations, whilst it changes under a $H_0 \setminus H_0 \cap H_1$ transformation in an opposite way to the way the bulk action does. The final form of the consistent partition function is then written as
\beq
Z^{\rm g.f.} \left[B^A\right] = \int \mathcal{D} B^i \; e^{i \Gamma_{\rm UV} \left[B^i\right]} \int \mathcal{D} \Sigma_1 \mathcal{D} A_\mu |^{\hat{A} = B^{\Sigma_1^{-1}}}_{\bar{A}} \; e^{i S_0 [A] + i \tilde{S}_{\rm CS}\left[A^\Lambda\right]}.
\eeq
This is invariant under any $H_0 \cup H_1$ transformations, although under an infinitesimal transformation $g \approx 1 - \hat{\gamma} \in G / (H_0 \cup H_1)$ it changes as
\beq
Z^{\rm g.f.} \left[ \left( B^a \right)^{\hat{\gamma}} \right] = \int \mathcal{D} B^i \; e^{i \Gamma_{\rm UV} \left[B^i\right]} \; e^{- ic \int_{\rm UV} \tilde{\omega}_4^{(1)} (\hat{\gamma}, B^i, B^a)} \int \mathcal{D} \Sigma_1 \mathcal{D} A_\mu |^{\hat{A} = B^{\Sigma_1^{-1}}}_{\bar{A}} \; e^{i S_0 [A] + i \tilde{S}_{\rm CS}\left[A^\Lambda\right]}.
\label{eq:SSB_gauging_general_transf}
\eeq 

We are now ready to discuss the dual 4D CFT description. The UV theory consists of the CFT sector and an external sector of Weyl fermions. The subgroup $H_0$ of $G$ is weakly gauged. 
In the IR, the CFT sector confines and this breaks $G$ down to a subgroup $H_1$. Unlike what happens in QCD, this unbroken subgroup $H_1$ does not need to be aligned with the gauged $H_0$, a typical situation occurring in dynamical symmetry breaking. We added proper local counter terms to the theory so that the UV-regulator preserves the $H_1$-symmetry. For this reason, the external fermions in the UV theory form an anomaly-free set under the $H_0 \cap H_1$ gauge forces. 

The anomaly story of this theory is rather rich: the anomaly phase in eq.~(\ref{eq:SSB_gauging_general_transf}) includes several different classes of anomalies. First, we have the gauge anomaly associated with the unbroken gauge symmetry $H_0 \cap H_1$. The UV CFT is free of this gauge anomaly since the UV-regulator respects the $H_0 \cap H_1$ symmetry (i.e~$\omega_5^{(0)} (A) \to \tilde{\omega}_5^{(0)} (A_{\mathbf{h}_1},A)$). For this reason, the external fermions do not carry any $H_0 \cap H_1$ anomalies. Second, the UV CFT has non-trivial anomalies in the global symmetry $G/H_0$. However, a subset $H_1\setminus H_0 \cap H_1 \subset G/H_0$ is anomaly free for the same reason why the gauged $H_0 \cap H_1$ is anomaly free. The remaining global symmetry $G/(H_0 \cup H_1)$ is, however, anomalous and the pure global anomaly or `t Hooft anomaly (e.g.~anomaly in $\langle J J J \rangle$ with all global currents $J$) is included in the phase in eq.~(\ref{eq:SSB_gauging_general_transf}). Third, the UV CFT has non-vanishing gauge anomalies for the broken part $H_0 \setminus H_0 \cap H_1$. This is cancelled by the chiral anomaly of the external fermions. We see that, generically, external fermions are charged under the gauged $H_0$, and while they carry non-trivial anomalies under the broken part $H_0 \setminus H_0 \cap H_1$, they form an anomaly-free set under $H_0 \cap H_1$. The anomaly phase in eq.~(\ref{eq:SSB_gauging_general_transf}) hence does not contain these anomalies. In fact, we need to look at the $\hat{\delta}$-variation to probe these, and the anomaly phase from the bulk (CFT) is exactly cancelled by that of $\Gamma_{\rm UV}$ (external fermions). There are several mixed anomalies we want to discuss. One of them is the mixed anomaly among gauged $H_0 \cap H_1$ and global $G/(H_0 \cup H_1)$ currents and is present in eq.~(\ref{eq:SSB_gauging_general_transf}). The local counter terms added to the theory result in the gauged current being conserved, and all the mixed anomaly is moved to the global currents. Next, there is a mixed anomaly among the gauged $H_0 \setminus H_0 \cap H_1$ and the global $G/(H_0 \cup H_1)$ currents, and eq.~(\ref{eq:SSB_gauging_general_transf}) includes these in principle. The requirement is that the addition of external fermions cancels any possible mixed anomaly in the gauged currents, and again all the mixed anomalies are stored in the global currents. 
In general, there may be also a mixed anomaly among gauged $H_0 \setminus H_0 \cap H_1$ and global $H_1 \setminus H_0 \cap H_1$, and we need to make sure that the mixed anomaly is all attributed to the global symmetry. This, and some of the above mentioned conditions, may require extra local counter terms on the UV brane. 

When the CFT sector confines in the IR, the gauge bosons associated with the broken generators acquire mass via the ``technicolor'' mechanism. Hence, these massive gauge bosons are integrated out in the IR, leaving only massless gauge bosons of $H_0 \cap H_1$. In addition, we have physical NGBs associated with $G/(H_0 \cup H_1)$. The ``would-be'' NGBs of $H_0 \setminus H_0 \cap H_1$ are eaten by the massive gauge bosons and removed in the unitary gauge. What happens to the external fermions? We recall that they couple to the broken gauge generators, and we think that they will also become massive and get integrated out. On the other hand, in the IR, we do not have any massless composite fermions in the spectrum, a dual picture of the absence of IR brane-localized fermions in the 5D. Hence, the anomalies of the UV theory described above must be matched by the WZW action made of the physical NGBs and the massless gauge bosons. Explicit confirmation of this deserves a separate designated study and we leave it for future works.

\subsubsection{Removing the ``would-be'' NGBs}

As already mentioned in the previous section, the NGBs associated with the broken \emph{gauged} generators are a gauge artifact and can be eliminated in the unitary gauge. In this section, we confirm that this expectation is fulfilled in our holographic description of anomaly inflow. In fact, we show that this is achieved via field redefinitions, or change of integration variables, which is an allowed operation for path-integrated, i.e.~dynamical, fields. Recall that the boundary value $B$ for the Dirichlet UV-BC corresponds to a classical source in the 4D dual theory. On the other hand, the boundary value $B$ for the Neumann UV-BC is dual to a dynamical gauge field and is path-integrated. Therefore, we see that for the latter case, the integration field variable can be redefined so as to eliminate the corresponding $\Sigma_1$-dependence. Such freedom, however, is absent for the purely global symmetry. In order to illustrate the point in a clean setup, instead of dealing with the most general case, in this section we consider the pure Neumann UV-BC.

Since all of the $G$ generators are gauged, none of $\Sigma_1$ corresponds to physical degrees of freedom. 
Therefore, we should be able to entirely remove the $\Sigma_1$-dependence. 
We first note that for small NGB field $\xi$, we can expand $\Sigma_1 \approx 1 - \xi$ and get
\beq
\begin{aligned}
\tilde{S}_{\rm CS} \left[A^\Lambda\right] &= c \int_{\rm 5D} \tilde{\omega}_5^{(0)} \left( \left( A^\Lambda \right)_h, A^\Lambda \right)  \\
& \approx   c \int_{\rm 5D} \tilde{\omega}_5^{(0)} (A_h, A) + c \int_{\rm 5D} d \tilde{\omega}_4^{(1)} (\xi, A_h, A).
\end{aligned}
\eeq
Using this, the partition function can be expressed as\footnote{Recall that $\Lambda$ is the 5D extension of the 4D $\Sigma_1$. Likewise, given a 4D $\hat{\xi}$ we extend it to 5D so that $\xi_{\rm UV} = \hat{\xi}$ and $\xi_{\rm IR} = 0$.}
\beq
Z^{\rm g.f.} = \int \mathcal{D} \Sigma_1 \mathcal{D} B \; e^{i \Gamma_{\rm UV} [B]} \; e^{- i c \int_{\rm UV} \tilde{\omega}_4^{(1)} \left(\hat{\xi}, B^{\Sigma_1^{-1}}\right)} \; \mathcal{D} A_\mu \vert^{\hat{A} = B^{\Sigma_1^{-1}}}_{\bar{A}} \; e^{i S_0 [A] + i \tilde{S}_{\rm CS} [A]}.
\eeq
The $\Sigma_1$-dependence can be eliminated by a change of variable, $B \to B^{\Sigma_1}$. Under this we get
%
%
%
%
%
%
\beq
\begin{aligned}
Z^{\rm g.f.} &= \int \mathcal{D} \Sigma_1 \mathcal{D} B^{\Sigma_1} \; e^{i \Gamma_{\rm UV} [B^{\Sigma_1}]} \; e^{- i c \int_{\rm UV} \tilde{\omega}_4^{(1)} \left(\hat{\xi}, B\right)} \; \mathcal{D} A_\mu \vert^{\hat{A} = B}_{\bar{A}} \; e^{i S_0 [A] + i \tilde{S}_{\rm CS} [A]} \\
&= \int \mathcal{D} \Sigma_1 \mathcal{D} B \; \left( e^{i G_{\rm UV} \left[\hat{\xi}, B\right]} \; e^{i \Gamma_{\rm UV} [B]} \right) \; e^{- i c \int_{\rm UV} \tilde{\omega}_4^1 \left(\hat{\xi}, B\right)} \; \mathcal{D} A_\mu \vert^{\hat{A} = B}_{\bar{A}} \; e^{i S_0 [A] + i \tilde{S}_{\rm CS} [A]}, 
\end{aligned}
\eeq
where $G_{\rm UV} \left[\hat{\xi}, B\right]$ is the anomaly functional of the external sector. 
Hence, we observe that the ``would-be'' NGB dependence is completely encoded in the two anomaly factors associated with the CFT and external degrees of freedom. In particular, we see that, provided the gauge anomaly cancels, the entire $\Sigma_1$-dependence disappears from the whole integrand, and the overall immaterial $\Sigma_1$-integral can be dropped. 
In other words, provided the gauged group is free of anomalies (hence suitable for gauging to begin with), the NGBs are unphysical and removable.

\section{Quantization conditions}
\label{sec:quantization}

The coefficient of the 5D CS action, called CS level, is subject to quantization conditions. 
A quick and easy way to derive this is by requiring that the variance of the 5D CS action under a gauge transformation is cancelled by the chiral anomaly of the 4D Weyl fermions localized on the boundaries. For simplicity we focus on a  $U(1)$ CS theory (see also~\cite{Hill:2006ei}). We first recall that the \emph{consistent} chiral anomaly\footnote{This is the form of the anomaly one gets from the triangle diagram. It is called \emph{consistent} anomaly since the result is consistent with Feynman diagram computations. In the presence of a single Weyl fermion, this is the only possible form. However, when the theory includes both LH and RH Weyl fermions, one can choose to decompose the currents in \emph{vector} and \emph{axial-vector}. In this case, there is an ambiguity in the form of the anomalous Ward identity. This ambiguity is due to possible local counter term(s) one can add to the theory. Equivalently, it is related to the UV regulator one chooses. A counter term of the form $S_{\rm CT} = \frac{1}{6\pi^2} \int d^4x \epsilon_{\mu\nu\rho\sigma} A^\mu V^\nu \partial^\rho V^\sigma$ can make the vector current conserved, while all of the anomaly is ``moved'' to the axial current:
\beq
\partial_\mu J^\mu_V = 0 , \qquad \quad \partial_\mu J^\mu_A = \frac{1}{8\pi^2} \left( F_V^{\mu\nu} \tilde{F}_{V \mu\nu} + \frac{1}{3} F_A^{\mu\nu} \tilde{F}_{A \mu\nu} \right).
\eeq  
This form of the anomaly is called \emph{covariant} anomaly since the theory is now invariant under the vector symmetry.
}
takes the form
\beq
\partial_\mu J^\mu_L = - \frac{1}{48 \pi^2} F_L^{\mu\nu} \tilde{F}_{L \mu\nu}, \qquad \quad%
\partial_\mu J^\mu_R = + \frac{1}{48 \pi^2} F_R^{\mu\nu} \tilde{F}_{R \mu\nu},
\eeq
with $J_L^\mu = \bar{\psi}_L \gamma^\mu \psi_L$ and $J_R^\mu = \bar{\psi}_R \gamma^\mu \psi_R$. When the Weyl fermions $\psi_L$ and $\psi_R$ have a vector-like coupling to the gauge field, $F_{L \mu\nu} = F_{R \mu\nu}$.
The main point here is that the chiral anomaly takes value in units of $\frac{1}{48\pi^2}$. The $U(1)$ CS theory in the bulk takes the form
\beq
S_{\rm CS} = c \int_{5D} A dA dA.
\eeq 
Under gauge transformations, we get
\beq
\delta S_{\rm CS} = \frac{c}{2} \left( \int_{\rm IR} d^4x \bar{\alpha} F \tilde{F} - \int_{\rm UV} d^4x \hat{\alpha} F \tilde{F} \right).
\eeq
Thus, the cancellation of these variance terms by $\kappa$ 4D Weyl fermions requires
\beq
c = \frac{\kappa}{24\pi^2}, \qquad \kappa \in \mathbb{Z}.
\eeq
For instance, for $\kappa>0$, the UV-localized variance term can be canceled by $\kappa$ RH fermions, while the IR-localized terms can be canceled by $\kappa$ LH fermions.

When the IR-BC is chosen so that the dual 4D picture describes spontaneous symmetry breaking, the relevant quantization condition comes from the argument by Witten in~\cite{Witten:1983tw} (see also~\cite{Panico:2007qd}). 
A proper reinterpretation of Witten's argument in the context of anomaly inflow may be given as follows.
First, we recall the instances in which we required the extension of the coset element from a map defined on 4D spacetime $S^4$ to the 5D bulk. For example, in the derivation of the gauge-fixed holographic effective action, we needed to make a change of variable with the property that $\tilde{\Lambda}_{\rm UV} = 1$ and $\tilde{\Lambda}_{\rm IR} = \Sigma_1 \in G/H_1$. Consequently, we extended the 4D $\Sigma_1 (x) \in G/H_1$ to a 5D object $\Lambda (x,z)$ such that $\Lambda_{\rm UV} = \Sigma_1 (x)$ and $\Lambda_{\rm IR} = 1$. Also, for any $g \in G$ transformation on the source field $B$, we assumed the existence of an extension to an element of the bulk gauge group $G_B$. By a decomposition $g(x) = \gamma(x) h (x), \gamma \in G/H_1, h \in H_1$, such an extension is equivalent to the possibility of deforming $\gamma (x)$ at the UV brane to $1$ (i.e.~$H_1$-element) at the IR brane. All these deformations/extensions have to do with the homotopy group $\pi_4 (G/H_1)$. When $\pi_4 (G/H_1) = 0$, we can extend $\Sigma_1 (x)$ to $D_5$. In what follows, we will assume this is the case.\footnote{As argued in~\cite{DHoker:1994rdl}, when $\pi_4 (G/H_1)$ is non-trivial, then we can deform $\Sigma_1 (x)$ to $\Sigma_a (x)$, a fixed representative of each homotopy class. Even in this case, the WZW action can still be defined with the same conclusions.}
Even when $\pi_4 (G/H) = 0$, there is an important question. Namely, one may consider two different extensions, $\Lambda_1 (x,z)$ and $\Lambda_2 (x,z)$, and the physics should be the same regardless of the extension chosen. Suppose $\Lambda_1$ is defined on $D_5$ and $\Lambda_2$ is defined on ${D'}_5$, with the spacetime manifold $S^4$ being the boundary of both $D_5$ and ${D'}_5$. 
The equivalence of the two extensions then imposes a quantization condition for the bulk CS action. 
Calling $S_{{\rm CS }i}$ the CS action with $\Lambda_i, \; i = 1,2$, the statement that the quantum action can be multi-valued only if the changes are integral multiples of $2\pi$ yields
\beq
e^{i \left( S_{\rm CS1} - S_{\rm CS2} \right) } = e^{i c \left( \int_{D_5} \omega_5^{(0)} - \int_{{D'}_5} \omega_5^{(0)} \right) } = e^{i c \int_{S^5} \omega_5^{(0)}} = e^{i 2\pi n}.
\eeq  
Therefore, if $\pi_5 (G/H_1)$ is trivial, then there is no quantization condition: all choices of $\Lambda (x,z)$ are (topologically) equivalent. If, on the other hand, $\pi_5 (G/H_1) \neq 0$, we are forced to choose the coefficient $c$ such that the above requirement is fulfilled. In the context of 4D QCD chiral symmetry breaking, $G= SU(3) \times SU(3)$ and $H_1 = SU(3)_V$, the fifth homotopy group is $\pi_5 (G/H_1) = \mathbb{Z}$ and Witten gave the correct normalization for the base 5-sphere~\cite{Witten:1983tw}.\footnote{Any other element of the $\pi_5 (G/H_1)$ is an integer multiple of the base sphere, hence once the integral over the base sphere is fixed, the remaining ones are just integer multiples of the value for the base sphere.} 

In the holographic description of anomaly inflow, we seem to have another way to determine the level $c$.
Namely, we may use the fact that a single parameter $c$ determines the anomaly of the UV and IR branes by the holographic realization of anomaly matching. Furthermore, changing UV and IR BC is equivalent to gauging (UV-BC) and spontaneous breaking (IR-BC), and it is conceivable that the ``norm'' of anomaly should not change under these ``deformations''.
For instance, given the bulk CS action, if we choose (partially) Neumann UV-BC, cancellation of the CS induced anomaly by the UV-localized Weyl fermions fixes the coefficient $c$. Then, if we choose IR-BC so that $G$ is broken to $H_1$, we get the WZW action with pre-determined coefficient. In fact, we used this to fix $c$ to be $c = \frac{\kappa}{24 \pi^2}$ in section~\ref{subsec:SSB_purely_global}, and successfully obtained the known normalization for the pure NGB WZW action. Furthermore, switching the UV-BC to a different one afterwards leads to a deformed theory with a fixed CS level.

\section{Conclusions}

In this paper, we systematically investigated the perturbative anomaly inflow by the bulk Chern-Simons (CS) theory in five-dimensional anti-de Sitter spacetime ($\textnormal{AdS}_5$). When the bulk geometry is $\textnormal{AdS}_5$, we are granted a holographic dual CFT description, in addition to the usual bulk--boundary interplay. The physics becomes especially rich and interesting once we introduce 3-branes. The UV brane allows us to incorporate ``weakly gauging'' of the global symmetry and as such the dual CFT possesses Adler, Bell, and Jackiw (ABJ) anomalies~\cite{Adler:1969gk, Bell:1969ts} as well as `t Hooft anomalies~\cite{tHooft:1980xss}. On the other hand, the phenomena of ``confinement'' and ``spontaneous symmetry breaking'' can be introduced once the IR brane is added. Therefore, through its dual description in terms of strongly coupled 4D CFT, anomaly inflow in the presence of UV and IR branes in $\textnormal{AdS}$ background captures a variety of features realized by fermion anomalies in quantum field theories.
A holographic approach to anomaly inflow turns out to provide an especially economic framework to organize and cover various facets of the fermion anomaly. Namely, one first makes a choice of bulk gauge group, based on the interest in Abelian vs non-Abelian anomaly, and anomaly of a single group vs mixed anomaly. Then, distinctions between the possible UV and IR BC integrate in a unified setup most, if not all, characteristics of chiral anomalies in confining gauge theories.

We showed that the choice of pure Neumann boundary conditions on the IR brane realizes `t Hooft anomaly matching holographically. 
We found that both ABJ (Neumann UV-BC) and `t Hooft (Dirichlet UV-BC) anomalies are matched by the composite fermions once 5D gauge invariance is restored by IR brane-localized 4D Weyl fermions. In the case of Neumann UV-BC, the necessary UV brane-localized fermions that restore 5D gauge invariance are physical spectator states. For Dirichlet UV-BC, as in the standard `t Hooft argument for the anomaly matching~\cite{tHooft:1980xss}, formally weakly gauging the global symmetry group is achieved by simply switching the Dirichlet UV-BC to Neumann. Once this change is made, then the existence of spectator fermions is required by 5D gauge invariance. Based on our findings that both ABJ and `t Hooft anomalies are matched by composite fermions when the symmetry $G$ is not broken on the IR brane, we infer that anomalies that inflow from the bulk CS theory are necessarily free of mixed anomalies with the confining gauge group of the 4D dual CFT. In the case of a mixed CS theory, taking $U(1) \times SU(2)$ as an example, we demonstrated that proper local counter terms can be added to the bulk CS action so that the mixed $U(1) \text{-} SU(2)$ anomaly is transferred between $U(1)$ and $SU(2)$ currents. We explicitly showed the form of local counter terms whose role is to attribute the entire mixed anomaly to either the $U(1)$ or $SU(2)$ currents.

We then worked out in detail the case of IR-BC such that the bulk gauge group $G$ is broken down to a subgroup $H_1$. 
This choice is dual to a spontaneously broken global symmetry of the CFT. 
By first considering pure Dirichlet UV-BC (purely \emph{global} symmetry of the CFT), we described how the Wess-Zumino-Witten action~\cite{Wess:1971yu, Witten:1983tw} naturally arises from the bulk CS action. In particular, we argued that unlike in the case of Neumann IR-BC, 5D gauge invariance does not require any IR brane-localized modes. Nevertheless, anomaly matching occurs via modes which are delocalized in the entire bulk. 
These modes, described as Wilson lines along the fifth direction, are indeed Goldstone bosons (GB) of spontaneously broken global symmetry, confirming our expectation based on the standard situation in QCD. When some part of $G$ is weakly gauged, by taking Neumann UV-BC, we discussed how our formulation leads to a unified description of both ABJ and `t Hooft anomalies. In particular, in such a case, we showed explicitly that the ``would-be'' GBs can be completely removed by means of field redefinitions. This matches the observation that the source fields for the weakly gauged symmetry become dynamical, and the associated GBs are eaten and become longitudinal modes of the gauge bosons of broken generators. In 5D, this is realized by the fact that the sources are path-integrated over once the UV-BC are chosen as Neumann, and this allows for the possibility of performing a change of variable. We also studied a very general case: $G$ broken down to $H_0$ by UV-BC and $G$ broken down to $H_1$ by IR-BC. This is a prototypical setup for models of dynamical symmetry breaking, in which a symmetry group $G$ is spontaneously broken to a subgroup $H_1$ and in addition another subgroup $H_0$ (not necessarily aligned with $H_1$) is weakly gauged.

Finally, we discussed the issue of quantization condition for the CS level. We described how such a requirement arises as an anomaly cancellation condition between brane-localized fermions and the bulk CS term when the BC is Neumann. On the other hand, when the BC is Dirichlet, we have reformulated Witten's argument~\cite{Witten:1983tw} in the context of anomaly inflow in $\textnormal{AdS}_5$ and shown that it indeed agrees with the condition obtained from Neumann BC. Since two distinct 4D theories with unbroken vs broken symmetry correspond to the same bulk theory in 5D with different BC, we speculated that the quantization condition of CS level might be insensitive to these ``deformations'' of the theory (i.e.~changes of BC). In this sense, our analysis opens up the possibility of fixing the level for one choice of BC, and then studying the theory with different BC and pre-determined normalization condition.



\section*{Acknowledgments}
We are grateful to Csaba Cs\'aki and Gowri Kurup for many helpful discussions and collaboration in the early stage of this work.
We also thank T.\ Daniel Brennan and Jay Hubisz for useful discussions.
S.H.\ would like to thank Minho Son for the invitation to present results of this work at ``KAIST Particle Theory Online Lecture Series''. S.H.\ is also grateful to the participants of ``KAIST Particle Theory Online Lecture Series'' for useful feedback and discussions.
G.R.\ thanks Cornell University for hospitality throughout the duration of this project.
S.H.\ is supported by the NSF grant PHY-2014071, and by Cornell University through the Hans Bethe Postdoctoral Fellowship. 
S.H.\ is also supported by a DOE grant DE-SC-0013642 and a DOE grant DE-AC02-06CH11357.
G.R.\ is supported in part by the DOE under grant award number DE-SC0009998.

\appendix
\section{Notation and conventions}
\label{appendix:notation_conventions}

Here we list the notation and conventions used throughout the paper. We work in Minkowski space with the mostly minus metric,
\begin{equation}
\eta_{\mu\nu}=\diag(1,-1,\dots,-1).
\end{equation}
We choose the generators of the gauge group to be anti-Hermitian, with commutation relations
\begin{equation}
\comm{T^a}{T^b}=f^{abc}T^c.
\end{equation}
A finite gauge transformation on the gauge field $A=A_\mu^aT^adx^\mu$ then results in
\begin{equation}
A^g=gDg^{-1}=g(d+A)g^{-1},
\end{equation}
where $D\equiv d+A$ is the covariant derivative. Under the same gauge transformation, the field strength $F=dA+A^2$ transforms as
\begin{equation}
F^g=gFg^{-1}.
\end{equation}
Writing $g=\exp(-\alpha)$ and expanding one can obtain the corresponding rules for infinitesimal gauge transformations:
\begin{equation}
\label{eq:infinitesimalgauge}
\delta_\alpha A=d\alpha+\comm{A}{\alpha},   \qquad   \quad%
\delta_\alpha F=\comm{F}{\alpha}.
\end{equation}
The Bianchi identity, which can be proved by using the definition of $F$ and properties of differential forms, reads
\beq
\label{eq:bianchi}
D F \equiv d F + \comm{A}{F} = 0.
\eeq

The generator of gauge transformations acting on functionals of the gauge field is
\begin{equation}
\label{eq:gaugegenerator}
X_a=-\partial_\mu\fdv{A_\mu^a}-f_{abc}A_\mu^b\fdv{A_\mu^c}=-D_\mu\fdv{A_\mu^a}.
\end{equation}
Thanks to this, we can construct the gauge transformation according to the following definition valid for any $\alpha$ and $F_a(x)$:
\begin{equation}
F_\alpha=\alpha\cdot F=\int d^nx\alpha^a(x)F_a(x).
\end{equation}
We can recover the transformation of eq.~(\ref{eq:infinitesimalgauge}) by acting with the generators on the gauge field:
\begin{equation}
\begin{aligned}
\delta_\alpha A&=X_\alpha A=\alpha\cdot XA=\int d^nx\alpha^a(x)X_a(x)A   \\
&=\int d^nx\alpha^a(x)\left(-\partial_\mu\fdv{A_\mu^a(x)}-f_{abc}A_\mu^b(x)\fdv{A_\mu^c(x)}\right)A_\nu^d(y)T^ddy^\nu   \\
&=\int d^nx\alpha^a\left(-\partial_\mu\delta^\mu_\nu\delta^d_a\delta(x-y)-f_{abc}A_\mu^b\delta^\mu_\nu\delta^d_c\delta(x-y)\right)T^ddy^\nu   \\
&=\int d^nx\left(\partial_\nu\alpha^d+f_{bad}A_\nu^b\alpha^a\right)T^ddy^\nu\delta(x-y)   \\
&=d\alpha+\comm{A}{\alpha}.
\end{aligned}
\end{equation}
In this sense, we can say that the covariant derivative can be ``integrated by parts'': in the first term because of actual integration by parts, in the second because of the totally antisymmetric nature of the structure constants.

We employ the following notation for both commutator and anticommutator:
\begin{equation}
\comm{A}{B}=AB-(-1)^{ab}BA,
\end{equation}
where $a$ and $b$ are 1 for anticommuting elements and 0 otherwise.

\section{Symmetry of the partition function}
\label{appendix:symmetry_of_Z_from_source_variation}

In this appendix, we review a standard fact about QFT: the symmetry property of the theory can be probed by checking how the partition function transforms as a functional of the source field. 
We first recall that even for a \emph{global} symmetry $G$, the Ward identity is derived by performing a \emph{local} version of the $G$ transformation. Since we are interested in studying gauge theories, we would like to consider 
\beq
Z [B_\mu] = \int \mathcal{D} \phi_i \; e^{i S [\phi_i] + i \int {\rm Tr} \left[ B_\mu^a J^{a \mu} \right]} = \left\langle e^{i \int {\rm Tr} \left[ B_\mu^a J^{a \mu}\right]} \right\rangle.
\eeq 
The $\phi_i$'s are fundamental fields of the underlying theory, and $J_\mu$ is the current they make up. 
Let's now look at the local $G$ transformation of $B_\mu$. Using 
\beq
g = e^{- \omega^a T^a} = e^{- \omega}, \qquad%
A_\mu = A_{\mu}^a T^a d x^\mu \to g \left( A +  d \right) g^\dagger \;\; \Rightarrow \;\; \delta A = d \omega + \left[ A , \omega \right] 
\eeq
we get
%
%
\beq
\begin{aligned}
Z \left[B^{g^{-1}}\right] &= \int \mathcal{D} \phi_i \; e^{i S[\phi_i] + i \int {\rm Tr} \left[ B^{g^{-1}} \cdot J \right]} \\
&= \int \mathcal{D} \phi_i \; e^{i S[\phi_i] + i \int {\rm Tr} \left[ B \cdot J^{g} + \omega d J\right]}. 
\end{aligned}
\eeq
Now, let us make a change of variable $\phi \to \phi^g$. Allowing for a possibly non-trivial (i.e.~anomalous) Jacobian factor but assuming that the action is invariant under this transformation, we get
\beq
\begin{aligned}
Z \left[B^{g^{-1}}\right] &=  \int \mathcal{D} \phi_i^g \; J \left[ \frac{\partial \phi}{\partial \phi^g} \right] \; e^{i S\left[\phi_i^g\right] + i \int {\rm Tr} [ B \cdot J^{g} + \omega d J]}  \\
&= \int \mathcal{D} \phi_i \; e^{-i \int {\rm Tr} [ \omega \mathcal{A} (B) ]} \; e^{i S [\phi_i] + i \int {\rm Tr} [ B \cdot J]} \; \left( 1 + i \int {\rm Tr} [ \omega d J ] \right),
\end{aligned}
\eeq
where we have written the anomalous Jacobian factor as an anomaly phase $e^{-i \int {\rm Tr} [ \omega \mathcal{A} (B) ]}$.
Therefore, we see that the condition that the partition function is invariant under a formal local transformation $B \to B^g = g (B+d) g^{-1}$ is equivalent to the statement that the global symmetry $G$ satisfies a complete set of (anomalous) Ward identities: $\langle d J \rangle = \mathcal{A}$. Furthermore, we can use this in a slightly different way. Namely, we instead perform a homogeneous transformation $B \to g B g^{-1}$. We see that under this, the partition function will change by $Z[B] \to Z [B^g] = e^{-i \int \omega \mathcal{A} (B)} Z [B]$. In other words, if the theory satisfies a non-anomalous Ward identity for $G$, we will not see this anomaly phase, while the anomaly phase will show up whenever the theory is actually anomalous under a $G$ transformation.

\section{Review of chiral anomalies}
\label{sec:review}



In this appendix, we review the standard treatment of fermion anomalies. After reviewing the Wess-Zumino consistency condition that the fermion anomaly should comply with, we introduce descent equations as a construction to solve the consistency condition. We then discuss the BRST formulation, and explain how chiral anomalies are defined as equivalence classes of BRST cohomology.

\subsection{Wess-Zumino consistency condition}
\label{subsec:WZ_consistency_condition}



Let $W$ be the generating functional for connected diagrams of a theory in the presence of external vector fields $A_\mu^a$ and other external (e.g.\ scalar, pseudo-scalar) fields. Up to contributions associated with those other external fields, the generators of a gauge transformation on $W$ are given by eq.~(\ref{eq:gaugegenerator}). The $X_a$'s fulfill the Lie algebra $\mathbf{g}$ of the gauge group $G$:
\beq
\left[ X_a (x), X_b (y) \right] = f_{abc} X_c (x) \delta (x-y).
\eeq
The anomalous Ward identity is expressed as
\beq
X_a (x) W \left[ A_\mu , \dots \right] = G_a (x).
\label{eq:anomalous_Ward_Id}
\eeq
Here $G_a (x)$ is the anomaly associated with $X_a$. The Wess-Zumino consistency condition follows by simply applying the Lie algebra to $W$:
\beq
X_a (x) G_b (y) - X_b (y) G_a (x) = f_{abc} G_c (x) \delta (x-y).
\label{eq:WZ_consistency_condition}
\eeq

This seemingly very simple condition has played a very important role in understanding deep aspects of anomalies. It made it possible to construct the form of the non-Abelian anomaly from pure geometrical considerations, without ever computing Feynman (triangle) diagrams. Along the way, it suggested a possible deep connection between anomalies in $D$ spacetime dimensions and higher-dimensional physics. It also allowed to successfully solve the anomalous Ward identity of eq.~(\ref{eq:anomalous_Ward_Id}), for known anomaly function $G(x)$, in the presence of Nambu-Goldstone bosons in the spectrum. It clarified the role of local counter terms in switching between consistent and covariant anomalies, and in ensuring that anomalies associated to an unbroken subgroup $H$ vanish.

\subsection{Descent equations and anomaly polynomial} 
\label{subsec:descent_eq}

The manipulations of geometrical objects that lead to the construction of the non-Abelian anomaly in $D=2n$ spacetime dimension are called \emph{descent equations}, and the resulting anomaly in its simplest form is sometimes dubbed as \emph{canonical anomaly}. As we will see, chiral anomalies are defined up to local counter terms (or equivalently up to choice of regulator). In order to make the discussion self-contained, we give a brief review of this idea. 

The goal is to construct an anomaly that satisfies the Wess-Zumino consistency condition. To be concrete, we assume that the $2n$-dimensional spacetime manifold is compact and closed (i.e.~there is no boundary). We first consider the following $(2n+2)$-dimensional form:
\beq
\Omega_{2n+2} (A) = {\rm Tr} \left( F^{n+1} \right),
\eeq
where $A$ is the gauge connection 1-form, and $F$ is its 2-form field strength. It is easy to show using the Bianchi identity eq.~(\ref{eq:bianchi}) that $\Omega_{2n+2}$ is a closed form:
\beq
d {\rm Tr} F^{n+1} = (n+1) {\rm Tr} \left( d F F^n \right) = - (n+1) {\rm Tr} \left( [A,F] F^n \right) = 0.
\eeq
Also, clearly $\Omega_{2n+2}$ is gauge invariant. Since it is closed, at least locally, we can write it as 
%
\beq
\Omega_{2n+2} (A) = d \omega_{2n+1}^{(0)} (A),
\eeq   
where the superscript ``0'' indicates that it is zeroth order in the gauge parameter.
We note, however, that the fully general solution is
\beq
\tilde{\omega}_{2n+1}^{(0)} (A) = \omega_{2n+1}^{(0)} (A) + d B_{2n} (A),
\eeq 
where the additional $2n$-form $B_{2n}$ is a local counter term, which is allowed simply because $d^2=0$. 
The gauge variation of $\tilde{\omega}_{2n+1}^{(0)}$ is closed. This is because $\Omega_{2n+2}$ is gauge invariant, and  
\beq
d \delta_\alpha \tilde{\omega}^{(0)}_{2n+1} = \delta_\alpha d \tilde{\omega}^{(0)}_{2n+1} = \delta_\alpha \Omega_{2n+2} = 0.
\eeq     
Therefore, at least locally, we can write 
\beq
\begin{gathered}
\delta_\alpha \tilde{\omega}_{2n+1}^{(0)} = d \tilde{\omega}_{2n}^{(1)} \qquad \textnormal{and} \qquad%
\delta_\alpha \omega_{2n+1}^{(0)} = d \omega_{2n}^{(1)},   \\
 \tilde{\omega}_{2n}^{(1)} (\alpha, A) = \omega_{2n}^{(1)} (\alpha, A) + \delta_\alpha B_{2n} (A) + d \hat{\omega}_{2n-1}^{(1)} (\alpha, A).
\end{gathered}
\eeq
Here the superscript ``1'' means the corresponding object is $\mathcal{O} \left(\alpha^1\right)$, and as before, in order to be fully general, we added the exact form $d \hat{\omega}_{2n-1}^{(1)} (\alpha, A)$. For spacetime manifolds without boundary, however, this term will not play any role, and we will drop it in the following.\footnote{As we will discuss momentarily, the anomaly that solves the Wess-Zumino consistency condition results from integrating $ \tilde{\omega}_{2n}^{(1)}(\alpha, A)$ over a manifold without boundary.}  
Let us now define two functionals. We consider the $(2n+1)$-dimensional disk $D_{2n+1}$, with $\partial D_{2n+1} = S^{2n}$:
\beq
\begin{gathered}
\tilde{W} [ A]  \equiv  \int_{D_{2n+1}} \tilde{\omega}_{2n+1}^{(0)} (A), \\
\tilde{G} [ \alpha, A ]  \equiv \int_{S^{2n}} \alpha^a (x) G_a (x) = \int_{S^{2n}} \tilde{\omega}_{2n}^{(1)} (\alpha, A).
\end{gathered}
\eeq
The claim is that $\tilde{G} [\alpha, A]$ is the anomaly consistent with the Wess-Zumino condition. To see this, first notice that
\beq
\delta_\alpha \tilde{W} [A] = \tilde{G} [ \alpha, A ].
\eeq   
In this sense, $\tilde{W} [A]$ is naively just like the effective action $W [A]$ we discussed above, but with the possible addition of local counter terms.\footnote{To be more precise, it is actually analogous to the Chern-Simons action in one more spacetime dimension, whose anomaly inflows to the boundary corresponding to the physical spacetime.} Then, since $\delta_{\alpha_1} \delta_{\alpha_2} - \delta_{\alpha_2} \delta_{\alpha_1} = \delta_{[\alpha_1, \alpha_2]}$,
\beq
 \delta_{\alpha_1} \tilde{G} [\alpha_2, A] - \delta_{\alpha_2} \tilde{G} [\alpha_2, A] = \tilde{G} [ [\alpha_1, \alpha_2],  A].
\eeq   
This shows that $\tilde{G} [\alpha, A]$ is indeed a functional of gauge fields which fulfills the Wess-Zumino consistency condition, and hence a candidate for the non-Abelian anomaly. 
We further note that 
\beq
\begin{gathered}
\tilde{G} [\alpha, A] = G [\alpha, A] + G_{\rm CT} [\alpha, A], \\
G [\alpha, A] = \int_{S^{2n}} \omega_{2n}^{(1)} (\alpha, A) \qquad \textnormal{and} \qquad%
G_{\rm CT} [\alpha, A] = \int_{S^{2n}} \delta_\alpha B_{2n} (A).
\end{gathered}
\eeq
One can repeat the above proof with $W [A] = \int_{D_{2n+1}} \omega_{2n+1}^{(0)} (A)$ to show that $G [\alpha, A]$ also solves the Wess-Zumino consistency condition. This $G [\alpha, A]$ is called the \emph{canonical anomaly}, and $G_{\rm CT} [\alpha,A]$ is the (shift in the) anomaly due to the addition of the local counter term $W_{\rm CT} [A] = \int_{D_{2n+1}} d B_{2n} (A)$. For obvious reasons, $\tilde{G} [\alpha, A]$ is then called the \emph{shifted anomaly}.

\subsection{BRST formalism}

In this appendix, we adopt the BRST formalism for chiral anomalies. 
An introduction to BRST formulation of chiral anomalies and descent equations can be found in~\cite{Weinberg:1996kr, Harvey:2005it} or in the original paper~\cite{Manes:1985df}.

\paragraph{Review of BRST invariance}
   
Let us begin the discussion with a very brief summary of BRST invariance. The path integral quantization of a gauge theory (with possibly matter fields $\psi$) results in
\beq
\mathcal{L} =  \mathcal{L}_{\rm matter} (\psi) - \frac{1}{4g^2} F_a^{\mu\nu} F_{a \mu\nu} - \partial_\mu \omega_a^* \partial^\mu \omega_a + f_{abc} (\partial_\mu \omega_a^*) A^\mu_c \omega_b + h_a f_a + \frac{1}{2} \xi h_a h_a,
\eeq
where $\omega$ ($\omega^*$) is an (anti-)ghost field and $h_a$ an auxiliary field that can be trivially integrated over in the path integral. A typical choice of gauge fixing is for example $f_a = \partial_\mu A^\mu_a$. This Lagrangian is not invariant under gauge transformations. It is, however, invariant under BRST transformations, with transformation parameter $\theta$ which anticommutes with ghost fields and all fermionic matter fields:
\beq
\begin{gathered}
\delta_\theta \psi = - t_a \theta \omega_a \psi, \qquad%
\delta_\theta A_{a\mu} = \theta D_\mu \omega_a,\qquad%
\delta_\theta \omega_a = - \frac{1}{2} f^{abc} \omega_b \omega_c, \\
\delta_\theta \omega^*_a = - \theta h_a, \qquad%
\delta_\theta h_a = 0.
\end{gathered}
\eeq
Notice that for matter and gauge fields, this is just a gauge transformation with transformation parameter $\theta \omega_a$. With proper assignment of ghost number (1 for ghost fields themselves, 0 for regular matter and gauge fields), one sees that a BRST transformation increases the ghost number by one. This motivates the introduction of the BRST operator $s$, whose action increases the ghost number by one, basically reproducing the above transformation rules. In terms of $s$ we get\footnote{Notice the notation used for the anticommutator of the Grassmann fields, as explained in appendix~\ref{appendix:notation_conventions}.}
\beq
\begin{gathered}
s A =  D(A) \omega, \qquad%
s \omega = - \frac{1}{2} [ \omega, \omega ], \\
s^2 = 0, \qquad%
\{ s , d \} = 0. 
\end{gathered}
\eeq
One says that the above set of equations (called structure equations) defines a graded commutative differential algebra. Here, $\omega$ plays the role of a 1-form along the ``ghost'' direction, and $s$ the ``exterior derivative'' along the ``ghost'' direction.

\paragraph{WZ consistency condition revisited}

In BRST language, the Wess-Zumino consistency condition is just the statement that the chiral anomaly is BRST-closed. To see why, let's define
\beq
G [\omega, A] = \int_{S^{2n}} \omega^a (x) G^a [x, A],
\eeq
where we replaced the gauge parameter with the ghost field $\omega (x)$. Then we get 
\beq
\begin{aligned}
s G [\omega, A] &= \int d^{2n} x \left( - \frac{1}{2} \comm{\omega (x)}{\omega (x)}^a \right) G^a [x, A] -\int d^{2n}x\omega^a(x)\left(\omega\cdot X\right)G^a[x,A] \\
&\hspace{-1cm}= -\frac{1}{2}\int d^{2n}xf^{abc}\omega^b(x)\omega^c(x)G^a[x,A]-\int d^{2n}xd^{2n}y\omega^a(x)\omega^b(y)X^b(y)G^a[x,A] \\
&\hspace{-1cm}=\frac{1}{2}\int d^{2n}xd^{2n}y\omega^a(x)\omega^b(y)\Bigl[X^a(x)G^b[y,A]-X^b(y)G^a[x,A]-f^{abc}G^c[x,A]\delta(x-y)\bigr].
\end{aligned}
\eeq
As expected, the Wess-Zumino consistency condition is indeed equivalent to the statement $s G[\omega, A] =0$: consistent chiral anomalies are BRST-closed. One of the immediate implications is that any addition of BRST-exact terms, i.e.~terms of the form $s B$, still result in a valid anomaly functional. Hence, chiral anomalies are equivalence classes of BRST cohomology~\cite{Manes:1985df}. This cohomology is defined by the operation $s$, and a BRST ``form'' takes as its coefficients local functionals of $A$ and its derivative $dA$ (or equivalently $A$ and $F$).

\paragraph{Descent equations revisited}

We can now repeat the descent equations procedure in BRST language:
\begin{itemize}
\item $\Omega_{2n+2} = d \omega_{2n+1}^{(0)}$ and $s \Omega_{2n+2} = 0$. 

Since $\Omega_{2n+2}$ has ghost number 0, the operator $s$ acts on it as a usual gauge transformation with $\omega$ as transformation parameter. 

\item $d \left( s \omega_{2n+1}^{(0)} \right) = - s d \omega_{2n+1}^{(0)} = - s \Omega_{2n+2} = 0$ implies that $s \omega_{2n+1}^{(0)} = d \omega_{2n}^{(1)}$.

Now the superscript denotes the degree in ghost number.

\item $d \left( s \omega_{2n}^{(1)} \right) = - s d \omega_{2n}^{(1)} = - s^2 \omega_{2n+1}^{(0)} = 0 $ implies that $s \omega_{2n}^{(1)} = d \omega_{2n-1}^{(2)}$.

In particular, this shows that $s \int_{S^{2n}} \omega_{2n}^{(1)} = \int_{S^{2n}} d \omega_{2n-1}^{(2)} = 0$ and, up to a numerical coefficient, $G \sim \int_{S^{2n}} \omega_{2n}^{(1)}$ is a consistent chiral anomaly. 
\end{itemize}
The entire discussion can be repeated if we replace $\omega$ with $\tilde{\omega}$, i.e.~with the addition of local counter terms $B_{2n}$. In fact, since the requirement for the consistent anomaly is $s G = 0$, shifting $\omega_{2n}^{(1)} \to \tilde{\omega}_{2n}^{(1)} = \omega_{2n}^{(1)} + s B_{2n}$ with an arbitrary $2n$-form still satisfies the condition. At the level of $\omega_{2n+1}^{(0)}$, this results in $\omega_{2n+1}^{(0)} \to \tilde{\omega}_{2n+1}^{(0)} = \omega_{2n+1}^{(0)} + d B_{2n}$, i.e.\ a shift in the Chern-Simons action.

\section{Cartan's homotopy formula and shifted anomaly}
\label{sec:Cartan_homotopy_formula}

\subsection{Cartan's homotopy formula}

In the previous appendix, we talked about the graded algebra defined by $d$ and $s$. Roughly, there are spacetime forms with corresponding exterior derivative $d$, and there are BRST forms with corresponding derivative $s$. In order to get a closed algebra, we required a relation between $d$ and $s$, which is given by $\{d ,s \} =0$. For Cartan's homotopy formula, we consider, instead of $s$ (or maybe \emph{in addition to} $s$), auxiliary ``directions'' defined by a family of continuous parameters $t = \{ t_1, t_2, \dots \}$. We want to construct the anti-derivation with respect to $t$, call it $d_t$, and establish a closed algebra between $d$ and $d_t$. It turns out that in addition to the \emph{odd} derivatives $d$ and $d_t$, we need an \emph{even} operator $\ell_t$. Now, we have a family of connections smoothly parametrized by $x$ and $t$, $A_t (x)$. $A_t (x)$ is a  $t$-form as well as an $x$-form. $d_t$ and $\ell_t$ are chosen such that the following graded algebra is satisfied:
\beq
\label{eq:commutator_lt_d}
\{ d, d\} = \{ d_t, d_t \} = \{ d, d_t \} = 0, \qquad%
[ \ell_t, d ] = d_t, \qquad%
[ d_t, \ell_t ] = 0.
\eeq
In a sense, the second equation, i.e.~$[ \ell_t, d ] = d_t$, is the only non-trivial one: it means $\ell_t$ decreases the degree in $dx$ by one, while at the same time increasing the degree in $dt$ by one. It provides a notion of ``$d^{-1}$'', which we need to determine the form of anomaly polynomials and local counter terms. It is not hard to show that for a polynomial of $\ell_t$, $f (\ell_t)$, we get
\beq
[ f (\ell_t) , d ] = d_t f' (\ell_t) = f' (\ell_t) d_t.
\label{eq:commutator_f_lt_d}
\eeq
The next step is to determine the action of $\ell_t$ on the algebra $\mathscr{Q}$ of polynomials generated by $\{ A_t, F_t \equiv d A_t + A_t^2, d_t A_t, d_t F_t \}$.\footnote{The reason we don't include $d F_t$ is that it is equivalent to $[ A_t, F_t]$ thanks to the Bianchi identity.} The unique action of $\ell_t$ on a polynomial $Q \in \mathscr{Q}$ is determined by requiring that (i) its action is consistent with the above graded algebra and (ii) the algebra $\mathscr{Q}$ is \emph{stable} under the action of $d, d_t$ and $\ell_t$. The resulting action of $\ell_t$ is
\beq
\ell_t A_t = 0, \qquad%
\ell_t F_t = d_t A_t .
\eeq
We are now ready to derive the \emph{differential form of the extended Cartan homotopy formula}. It is obtained by applying eq.~(\ref{eq:commutator_f_lt_d}) to $f (\ell_t) = \ell_t^{p+1}/(p+1)!$. For a polynomial $Q \in \mathscr{Q}$, we get
\beq
d_t \frac{\ell_t^p}{p!} Q = \frac{\ell_t^{p+1}}{(p+1)!} d Q - d \frac{\ell_t^{p+1}}{(p+1)!} Q.
\label{eq:Cartan_homotopy_formula_diff}
\eeq
A more useful formula for the study of anomalies is obtained by integrating eq.~(\ref{eq:Cartan_homotopy_formula_diff}) over the domain $T$ of $t$. In order to do this, we need to establish a convention for the incomplete integration. We follow~\cite{Manes:1985df} for the conventions: if $\alpha$ is a form of degree $(r,s)$ in $(dx, dt)$, we have
\beq
\begin{gathered}
\int_{X} \alpha \equiv \int_{X} \alpha_{r,s} dx^r dt^s = \left( \int_{X} \alpha_{r,s} dx^r \right) dt^s, \\
\int_{T} \alpha \equiv \int_{T} \alpha_{r,s} dx^r dt^s = (-1)^{rs} \left( \int_{T} \alpha_{r,s} dt^s \right) dx^r.
\end{gathered}
\eeq
It follows that 
\beq
\begin{gathered}
d \int_{T} \alpha = (-1)^s \int_{T} d \alpha, \qquad%
d_t \int_{X} \alpha = (-1)^r \int_{X} d_t \alpha, \\
\int_{X} d \alpha = \int_{\partial X} \alpha, \qquad%
\int_{T} d_t \alpha = \int_{\partial T} \alpha. 
\end{gathered}
\eeq
While it can be explicitly checked, the first two equations imply that $\int_{X}$ $\left(\int_{T}\right)$ behaves like an odd operation of degree $r$ ($s$). The last two show that Stokes' theorem works as usual. The \emph{integral form of the extended Cartan homotopy formula} is obtained by integrating eq.~(\ref{eq:Cartan_homotopy_formula_diff}): 
\beq
\int_{\partial T} \frac{\ell_t^p}{p!} Q = \int_T \frac{\ell_t^{p+1}}{(p+1)!} d Q - (-1)^{p+q+1} d \int_T \frac{\ell_t^{p+1}}{(p+1)!} Q,
\eeq 
where $q$ is the degree of $Q$ in $dt$. Notice that $\ell_t^{p+1} Q/(p+1)!$ has degree $(p+q+1)$ in $dt$, and therefore so does $\int_T$.

\subsection{Derivation of the shifted anomaly and local counter terms}
\label{subsec:shifted_anomaly_local_counter_term}

So far Cartan's homotopy formula has been introduced as a pure mathematical expression for a given algebra of $d, d_t$ and $\ell_t$. There are three independent choices one can make when the formula is used: $p,q$, and the choice of parametrization for $A_t$. Now remember that $\ell_t$ provides an operation that behaves like a ``$d^{-1}$''. From the descent equations, we learn that $\Omega_{2n+2} = d \omega_{2n+1}^{(0)}$, and $s \omega_{2n+1}^{(0)} = d \omega_{2n}^{(1)}$. While those tell us that objects like $\omega_{2n+1}^{(0)}$ and $\omega_{2n}^{(1)}$ exist, we still need a way to determine their explicit form. There is no \emph{unique} inverse of $d$ to simply solve these equations. Different choices of ``$d^{-1}$'' will lead to different $\omega_{2n+1}^{(0)}$ and $\omega_{2n}^{(1)}$, related to each other by the ambiguity associated with local counter terms. $\ell_t$ provides one such operation. Importantly, depending on the choice of $A_t$ (and in principle also $p$, although we will exclusively take $p=0$), we can get different answers.

We first mention a few possible options for $A_t$:
\beq
\begin{gathered}
\label{eq:cartan_param}
A_t = t A, \;\; t \in [0,1], \qquad \qquad%
A_t = t A_1 + (1-t) A_0, \;\; t \in [0,1], \\
A_t = t_1 A_1 + t_2 A_2 + (1-t_1-t_2) A_0, \;\; t_1,t_2 \in [0,1].
\end{gathered}
\eeq
The first connects $A$ to nothing between $t=0$ and $t=1$. It will turn out to be a useful choice for a theory with simple symmetry group $G$ without any symmetry breaking. The second connects two different configurations $A_0$ and $A_1$. We will find it useful when we need to distinguish a particular part of the group (or algebra), notably in the case of $G/H$ or a product group $G_1 \times G_2$. The last is an example with more than one $t$, and was used in~\cite{Manes:1985df} to derive the ``triangle formula'' and other results. In order to avoid confusion, in the following we will refer to the parameters of the first and second parametrization as $t$ and $s$, respectively.\footnote{Notice that $s$ here is not the BRST operation. We will never encounter both in one place.} Let us now consider some examples:
\begin{itemize}
\item[(1)] $Q_t = \Omega_{2n+2,t} = {\rm Tr} (F_t^{n+1}), \; A_t = t A, \; p=q=0$.

Cartan's homotopy formula (in its integrated version) with these choices results in
\beq
\Omega_{2n+2,1} - \Omega_{2n+2,0} = \Omega_{2n+2} (A) = d \int_T \ell_t \Omega_{2n+2,t} \equiv d \omega_{2n+1}^{(0)},
\label{eq:Cartan_eq_first_param}
\eeq
where we used the subscript $t=0,1$ to denote that the corresponding functional is in terms of $A_t$ with a specified $t$-value. An explicit expression for $\omega_{2n+1}^{(0)}$ is then given by
\beq
\omega_{2n+1}^{(0)} = \int_T \ell_t \Omega_{2n+2,t} = (n+1) \int_0^1 dt \; {\rm Tr} \left[ A F_t^n \right].
\label{eq:canonical_omega_2n+1}
\eeq
This expression was indeed obtained in~\cite{Zumino:1983rz} with a different method. According to the language used in~\cite{Chu:1996fr}, the chiral anomaly obtained from this is the \emph{canonical anomaly}.
\item[(2)] $Q_s = \Omega_{2n+2,s}, \; A_s = s A_1 + (1-s) A_0, \; p=q=0$.

This time the integrated form of Cartan's homotopy formula gives
\beq
\Omega_{2n+2,1} - \Omega_{2n+2,0} = d \int_S \ell_s \Omega_{2n+2,s} \equiv d \tilde{\omega}_{2n+1}^{(0)}.
\label{eq:Cartan_eq_second_param}
\eeq
Explicitly, $\tilde{\omega}_{2n+1}^{(0)}$ is 
\beq
\tilde{\omega}_{2n+1}^{(0)} (A_0,A_1) = (n+1) \int_0^1 ds \; {\rm Tr} \left[ (A_1 - A_0) F_s^n \right].
\label{eq:shifted_omega_2n+1}
\eeq
\item[(3)] Relation between $\omega_{2n+1}^{(0)}$ and $\tilde{\omega}_{2n+1}^{(0)}$.

From (1), we know that each $\Omega_{2n+2,i}, \; i=0,1$ appearing on the LHS of eq.~(\ref{eq:Cartan_eq_second_param}) can be expressed as $d \omega_{2n+1,i}^{(0)}$ by eq.~(\ref{eq:Cartan_eq_first_param}). Roughly, we get
\beq
\label{eq:cartancounter term}
d \omega_{2n+1,1}^{(0)} - d \omega_{2n+1,0}^{(0)} = d \tilde{\omega}_{2n+1}^{(0)} \quad \Rightarrow \quad \tilde{\omega}_{2n+1}^{(0)} = \omega_{2n+1,1}^{(0)} - \omega_{2n+1,0}^{(0)} + d B_{2n},
\eeq 
where the total derivative accounts for the presence of the local counter term. How do we determine the form of $B_{2n}$? First of all, we define $\omega_{2n+1,s}^{(0)}$ in the following way for $i=0,1$:
\beq
\Omega_{2n+2,i} = d \omega_{2n+1,i}^{(0)}, \qquad \omega_{2n+1,i}^{(0)} = \int_T \ell_t \Omega_{2n+2,it}, \qquad A_{it}=t A_i.
\eeq
Then we note that $A_0$ and $A_1$ can be homotopically connected by extending $i$ to $s$, which results in the following parametrization:
\beq
A_{st} = t A_s = t \left( s A_1 + (1-s) A_0 \right), \qquad s\in [0,1].
\eeq
In particular, this means that we now have an object $\omega_{2n+1,s}^{(0)}$ given by
\beq
\omega_{2n+1,s}^{(0)} = \int_T \ell_t \Omega_{2n+2,st}, \;\; A_{st}=t A_s = t \left( s A_1 + (1-s) A_0 \right),
\eeq
which makes it possible to write down Cartan's homotopy formula for $\omega_{2n+1,s}^{(0)}$. It is given by
\beq
\begin{aligned}
\omega_{2n+1,1}^{(0)} - \omega_{2n+1,0}^{(0)} &= \int_S \ell_s d \omega_{2n+1,s}^{(0)} + d \int_S \ell_s \omega_{2n+1,s}^{(0)} \\
&= \int_S \ell_s \Omega_{2n+2,s} + d \int_S \ell_s \omega_{2n+1,s}^{(0)} \\
&= \tilde{\omega}_{2n+1}^{(0)} + d \int_S \ell_s \omega_{2n+1,s}^{(0)}.
\label{eq:shifted_omega_canonical_omega_and_B_2n}
\end{aligned}
\eeq
Comparing with eq.~(\ref{eq:cartancounter term}), this yields the result
\beq
B_{2n} = - \int_S \ell_s \omega_{2n+1,s}^{(0)}. 
\label{eq:B_2n}
\eeq
The expressions for the counter term and the related shifted Chern-Simons action indeed agree with the ones given in~\cite{Chu:1996fr}.
\item[(4)] Explicit expression for $B_{2n}$.

An explicit expression for $B_{2n}$ is obtained by evaluating the integral, and was done in~\cite{Chu:1996fr}. So, instead of repeating the computation, we will simply quote the result for $n=2$. $B_4$ is given by  
\beq
B_4 (A_h, A) = \frac{1}{2} {\rm Tr} \left[ \left( A_h A - A A_h \right) \left( F + F_h \right) + A A_h^3 - A_h A^3 + \frac{1}{2} A_h A A_h A \right],
\label{eq:B_4}
\eeq
where $F_h = d A_h + A_h^2$. 
\item[(5)] Useful properties.

The shifted CS action $\tilde{\omega}_{2n+1}^{(0)}$ and the local counter term $d B_{2n}$ satisfy the following properties, which are used frequently throughout the paper and follow from eq.~(\ref{eq:shifted_omega_2n+1}) and (\ref{eq:shifted_omega_canonical_omega_and_B_2n}):
\begin{itemize}
\item[\circled{1}] $\tilde{\omega}_{2n+1}^{(0)} (A_0^g, A_1^g) = \tilde{\omega}_{2n+1}^{(0)} (A_0, A_1)$.

The shifted CS action is invariant under \emph{simultaneous} gauge transformations of $A_0$ and $A_1$, $A_i \to g \left( A_i + d \right) g^{-1}$. We emphasize that this holds regardless of the relation~(\ref{eq:shifted_omega_canonical_omega_and_B_2n}) and, in particular, of whether $\omega_{2n+1,0}^{(0)}$ vanishes (as in the $G/H$ case) or not (as with a general $G_1 \times G_2$). This is simply because, under simultaneous transformations, $(A_1 - A_0) \to g (A_1 - A_0) g^{-1}$, i.e.~the derivative terms cancel and the whole term transforms as adjoint matter, so we can use cyclicity of the trace in eq.~(\ref{eq:shifted_omega_2n+1}).
\item[\circled{2}] $\tilde{\omega}_{2n+1}^{(0)} (A_0, A_1 = A_0 ) = 0$.
\item[\circled{3}] $d B_{2n} (A_0, A_1 = A_0) = 0$.
\item[\circled{4}] $\tilde{\omega}_{2n+1}^{(0)} (A_0=0, A_1) = \omega_{2n+1}^{(0)} (A_1)$.

To see this, we note that with $A_0 = 0$, we get $A_t \to t A_1$, and eq.~(\ref{eq:shifted_omega_2n+1}) reduces to 
\beq
\tilde{\omega}_{2n+1}^{(0)} (A_0=0, A_1) = (n+1) \int_0^1 dt \; {\rm Tr} \left[ A_1 F_t^n \right] = \omega_{2n+1}^{(0)} (A_1).
\eeq
\item[\circled{5}] $d B_{2n} (A_0=0, A_1) = 0$.
\item[\circled{6}] $\tilde{\omega}_{2n+1}^{(0)} (A_0, A_1=0) = - \omega_{2n+1}{(0)} (A_0)$.
\item[\circled{7}] $d B_{2n} (A_0, A_1=0) = 0$.

\end{itemize} 
The first property leads to many interesting and important consequences. We discuss a few of them below.
\item[(6)] $G/H$ symmetry group.

We consider a symmetric space. Let's denote the algebra for the unbroken group $H$ as $\mathbf{h}$, and that of the coset as $\mathbf{k}$. Referring to the corresponding elements as $h \in \mathbf{h}, \; k \in \mathbf{k}$, we want
\beq
[ h, h ] \subseteq \mathbf{h} \;\; ({\rm subalgebra}), \quad%
[ h, k ] \subseteq \mathbf{k} \;\; ({\rm reductive}), \quad%
[ k, k ] \subseteq \mathbf{h} \;\; ({\rm symmetric}).
\eeq
The gauge field $A$ and the (infinitesimal) group element $\alpha \in \mathbf{g}$ can be decomposed into their $\mathbf{h}$ and $\mathbf{k}$ parts:
\beq
\begin{gathered}
A = A_h + A_k, \qquad\alpha = \beta + \gamma,  \\
A_h, \beta \in \mathbf{h}, \qquad A_k, \gamma \in \mathbf{k}. 
\end{gathered}
\eeq
Under an infinitesimal gauge transformation,
\beq
\delta_\alpha A =  \underbrace{\left( d \beta + [ A_h, \beta ] + [ A_k , \gamma ] \right)}_{= \; \delta_\alpha A_h}
+ \underbrace{\left( d \gamma + [ A_h, \gamma ] + [ A_k , \beta] \right)}_{= \; \delta_\alpha A_k}.
\eeq
For a general $\alpha \in \mathbf{g}$, $A$ and $A_h$ do not transform in the same way. They do so, however, under $\beta \in \mathbf{h}$. Therefore, if we set $A_0 = A_h$ and $A_1 = A$, the resulting shifted Chern-Simons action is invariant under $H$. Or, equivalently, the modified effective action is invariant under any $H$-transformation: for all $\beta \in \mathbf{h}$, $G_\beta = \delta_\beta W = 0$. This is indeed a crucial assumption in the derivation of the original Wess-Zumino action. The current discussion allows us to determine what local counter term needs to be added. If, in addition, $H$ is an anomaly-free embedding (AFE), then $\omega_{2n+1,0}^{(0)} = \omega_{2n+1}^{(0)} (A_h) = 0$ and
\beq
\tilde{\omega}_{2n+1}^{(0)} (A_h, A) = \omega_{2n+1}^{(0)} (A) + d B_{2n} (A_h, A).
\eeq
\item[(7)] Product group $G_1 \times G_2$ and mixed anomaly.

Let's comment on the case in which $G_1 = U(1)$ and $G_2$ is non-Abelian (e.g.~$SU(N)$). In this example we have a mixed anomaly, and the role of the local counter term is to attribute the anomaly to one of the following: the $G_1$ current, the $G_2$ current, or a combination of the two. This is exactly what the counter term $B_{2n}$ constructed above does, as it ensures that the anomaly associated with a chosen subgroup vanishes. To see this, we first formally embed $G_1 \times G_2$ into a larger group (e.g.~with $G_1$ and $G_2$ as diagonal blocks), so that we can formally write gauge fields and (infinitesimal) group elements as 
\beq
\begin{gathered}
A = A_1 + A_2, \qquad A_1 \in \mathscr{G}_1, \; A_2 \in \mathscr{G}_2, \\
\alpha = \alpha_1 + \alpha_2, \qquad \alpha_1 \in \mathscr{G}_1, \; \alpha_2 \in \mathscr{G}_2.
\end{gathered}
\eeq
Then under gauge transformations of each factor group, we get
\beq
\begin{gathered}
({\rm Under} \; \alpha_1 \in \mathscr{G}_1) \quad A \to A + d \alpha_1, \quad A_1 \to A_1 + d \alpha_1, \\
({\rm Under} \; \alpha_2 \in \mathscr{G}_2) \quad A \to A + d \alpha_2 + [ A, \alpha_2 ], \quad A_2 \to A_2 + d \alpha_2 + [ A_2, \alpha_2 ].
\end{gathered}
\eeq
In the second line, we used the fact that $\mathscr{G} = \mathscr{G}_1 \oplus \mathscr{G}_2$, i.e.~$[A_1, \alpha_2] = 0$. Therefore, we see for example that if we take $A_0 \equiv A_2$ and $A_1 \equiv A$ in the sense of the parametrization of eq.~\ref{eq:cartan_param}, then those fields transform under $\alpha_2 \in \mathscr{G}_2$ in the same way, and $\tilde{\omega}_{2n+1} (A_0, A_1)$ will be invariant under a $G_2$-transformation. In other words, the associated counter term moves the entire anomaly to the $U(1)$ current. The opposite works just as well: if we instead take $A_0 \equiv A_1$ and $A_1\equiv A$, then this time all the mixed anomaly is attributed to the $G_2$-current. Our discussion is very general, and these statements hold for any choice of $G_2$ that has a mixed anomaly with $G_1=U(1)$.

For the sake of completeness, we briefly comment on the algebra. While taking $\mathscr{G}_2 = \mathbf{h}$ and $\mathscr{G}_1 = \mathbf{k}$ results in the algebra of a symmetric space, this is not the case if we instead take $\mathscr{G}_1 = \mathbf{h}$ and $\mathscr{G}_2 = \mathbf{k}$. This may have potential implications for WZW terms.

\end{itemize}

\section{Wess-Zumino-Witten action}
\label{sec:WZW_review}

In this appendix, we first review the motivation for the Wess-Zumino-Witten (WZW) action~\cite{Wess:1971yu,Witten:1983tw} following the original argument by Wess and Zumino~\cite{Wess:1971yu}. 
We then present a construction for the WZW action in terms of the shifted CS action $\tilde{\omega}^{(0)}_{2n+1}$~\cite{Chu:1996fr}.

Suppose we have a theory with effective action (for connected diagrams) $W [V_\mu, \xi, \dots]$ which consists of (external) vector fields ($V_\mu$), (external) (pseudo-)scalars $\xi$, and some other fields. We further assume that the symmetry group $G$ is spontaneously broken down to $H$ and the theory is anomalous under the gauge transformation:
\beq
\delta_\alpha W [V_h,V_k, \xi] = \left( X_i \beta_i + Y_a \gamma_a + Z_A \alpha_A + \dots \right) W [V_h,V_k, \xi, \dots] = G_\alpha [V_h,V_k],
\label{eq:anomalous_Ward_id_for_WZW}
\eeq
where we split the gauge fields into their $\mathbf{h}$ and $\mathbf{k}$ parts, $V = V_h + V_k$. We also do the same for the transformation parameter, $\alpha = \beta + \gamma$, with $\beta \in \mathbf{h}$ and $\gamma \in \mathbf{k}$. Accordingly, $X_i$ is the generator associated with $V_h$, $Y_a$ with $V_k$, and $Z_A$ with the (pseudo-)scalars $\xi$. The associated anomaly functional $G_\alpha$ is a functional of gauge fields only. In perturbation theory, we know the effective action $W$ and can compute its variation, hence the anomaly, by computing Feynman diagrams. The situation is much less clear in strongly coupled theories, especially when the theory confines, and generically $W$ is not known in the confined phase. If, on the other hand, one knows the anomaly functional $G_\alpha$ (e.g.~by computing it in the quark-gluon phase and using RG invariance of the chiral anomaly), one may hope to solve the above anomalous Ward identity to get the effective action $W$ in the strongly coupled IR phase. Mathematically, the question Wess and Zumino addressed in their work~\cite{Wess:1971yu} is whether we can solve for $W$ once $G_\alpha$ is known.

For an explicit construction of the WZW action, we first consider a spacetime manifold $S^{2n}$. We assume that the gauge field $B$ on $S^{2n}$ can be extended to $A$ on the disk $D_{2n+1}$, whose boundary is $S^{2n}$. The boundary value of the gauge field on $D_{2n+1}$ is the gauge field we started with on $S^{2n}$: $A \vert_{S^{2n}} = B$. Next, we consider $\Sigma = e^{- \xi} \in G/H$. As we did for the gauge field, we need to extend this to $D_{2n+1}$ as well: $\Sigma \to \Lambda$ in such a way that $\Lambda \vert_{S^{2n}} = \Sigma$. This is possible if $\pi_{2n} (G/H)$ is trivial.

Now, the claim is that the WZW effective action can be written in terms of the shifted CS action as
\beq
W [\Sigma , B] = \int_{D_{2n+1}} \left[\tilde{\omega}_{2n+1}^{(0)} (A_h, A) - \tilde{\omega}_{2n+1}^{(0)} \left(\left(A^{\Lambda^{-1}}\right)_h, A^{\Lambda^{-1}}\right)\right].
\label{eq:WZW_action_in_terms_of_tilde_omega}
\eeq
In order to confirm this, it suffices to check the following:
\begin{itemize}
\item[\circled{1}] Eq.~(\ref{eq:WZW_action_in_terms_of_tilde_omega}) solves the anomalous Ward identity. 

To see this, we first show that $\tilde{\omega}_{2n+1}^{(0)} \left(\left(A^{\Lambda^{-1}}\right)_h, A^{\Lambda^{-1}}\right)$ is \emph{invariant} under any $g \in G$ transformation. 
Indeed we have
\beq
A^{\Lambda^{-1}} \to \left( A^g \right)^{(\Lambda^g)^{-1}} = A^{h \circ \Lambda^{-1} \circ g^{-1} \circ g} = h(\Lambda, g) \left( d + A^{\Lambda^{-1}} \right) h^{-1} (\Lambda, g).
\eeq
In other words, under an arbitrary $g \in G$, the combination $A^{\Lambda^{-1}}$ transforms under a particular $H$-element defined by $g \Lambda = \Lambda^g h \left( \Lambda, g \right)$ as if it were a single gauge field. This proves the statement since while $\tilde{\omega}_{2n+1}^{(0)} (A_h, A)$ is not invariant under an arbitrary $G$ transformation, it is invariant under any $H$ transformation.
This also implies that any non-trivial anomaly only comes from the first term in eq.~(\ref{eq:WZW_action_in_terms_of_tilde_omega}), which is to be expected since the usual shifted CS term solves the anomalous Ward identity (with shifted anomaly).
\item[\circled{2}] It vanishes as $\Sigma \to 1$.

This is easily seen by noting that $\Lambda \to 1$ as $\Sigma \to 1$. Hence, there is no solution to the anomalous Ward identity which is a function of gauge fields only.
\item[\circled{3}] It is a function of the \emph{boundary} values, $\Sigma$ and $B$.

It is sufficient here to show that the integrand is a closed form:
\beq
d \left( \tilde{\omega}_{2n+1}^{(0)} (A_h, A) - \tilde{\omega}_{2n+1}^{(0)} \left(\left(A^{\Lambda^{-1}}\right)_h, A^{\Lambda^{-1}}\right) \right) = \Omega_{2n+2} (A) - \Omega_{2n+2} \left(A^{\Lambda^{-1}}\right) = 0,
\eeq
as $\Omega_{2n+2}$ is gauge invariant. Therefore, at least locally, the integrand can be written as an exact $2n$-form, and the integral over the disk depends only on the boundary values, $\Sigma$ and $B$.
\end{itemize}

\bibliographystyle{JHEP}
\bibliography{refs}

\end{document}